\documentclass[useAMS,usenatbib]{mn2e}
\usepackage{epsfig}

\def\gsim{\;\lower4pt\hbox{${\buildrel\displaystyle >\over\sim}$}\;}
\def\lsim{\;\lower4pt\hbox{${\buildrel\displaystyle <\over\sim}$}\;}
\def\grls{\;\lower4pt\hbox{${\buildrel\displaystyle >\over <}$}\;}
\title[Similarity Isothermal MHD Shocks]
{Envelope Expansion with Core Collapse. III.\\
Similarity Isothermal Shocks in a Magnetofluid }
\author[Yu, Lou, Bian, Wu]{Cong Yu$^{1, 5}$,
\thanks{E-mail: louyq@tsinghua.edu.cn; lou@oddjob.uchicago.edu;
yccit@yahoo.com.cn}
Yu-Qing Lou$^{2,3,4}$, Fu-Yan Bian$^{2}$, and Yan Wu$^{2}$\\
$^{1}$National Astronomical Observatories/Yunnan Astronomical
Observatory, Chinese Academy of Sciences, Kunming, 650011, China;\\
$^{2}$Physics Department and Tsinghua Centre for Astrophysics
(THCA), Tsinghua University, Beijing, 100084, China;\\
$^{3}$Department of Astronomy and Astrophysics, the University
of Chicago, 5640 South Ellis Avenue, Chicago, IL 60637, USA;\\
$^{4}$National Astronomical Observatories of China, Chinese
Academy of Sciences, A20, Datun Road, Beijing 100012, China;\\
$^{5}$Graduate School of the Chinese Academy of Sciences, Beijing, China.}
\begin{document}

\date{Accepted 2006 ?? ??. Received 2006 ?? ??;
in original form 2005 ?? ??}

\pagerange{\pageref{firstpage}--\pageref{lastpage}} \pubyear{2004}

\maketitle

\label{firstpage}

\begin{abstract}
We explore magnetohydrodynamic (MHD) solutions for envelope expansions
with core collapse (EECC) with isothermal MHD shocks in a quasi-spherical
symmetry and outline potential astrophysical applications of such
magnetized shock flows. By including a random magnetic field in a gas
medium, we further extend the recent isothermal shock results of Bian
\& Lou who have unified earlier similarity isothermal shock solutions
of Tsai \& Hsu, of Shu et al. and of Shen \& Lou in a more general
framework. MHD shock solutions are classified into three classes
according to the downstream characteristics near the core. Class I
solutions are those characterized by free-fall collapses towards the
core downstream of an MHD shock, while Class II solutions are those
characterized by Larson-Penston (LP) type near the core downstream of
an MHD shock. Class III solutions are novel, sharing both features of
Class I and II solutions with the presence of a sufficiently strong
magnetic field as a prerequisite. Various MHD processes may occur
within the regime of these isothermal MHD shock similarity solutions,
such as sub-magnetosonic oscillations, free-fall core collapses,
radial contractions and expansions. Both possibilities of
perpendicular and oblique MHD shocks are analyzed. Under the current
approximation of MHD EECC solutions, only perpendicular shocks are
systematically calculated. These similarity MHD shocks propagate at
either sub-magnetosonic or super-magnetosonic constant speeds. We can
construct Class I, II and III MHD shocks matching with an isothermal
magnetostatic outer envelope or an MHD breeze. We can also construct
families of twin MHD shock solutions as well as an `isothermal MHD
shock' separating two magnetofluid regions of two different yet constant
temperatures. The versatile behaviours of such MHD shock solutions may
be utilized to model a wide range of astrophysical problems, including
star formation in magnetized molecular clouds, `MHD champagne flows'
in HII regions around luminous massive OB stars, MHD link between the
asymptotic giant branch phase to the proto-planetary nebula phase with
a hot central magnetized white dwarf, relativistic MHD pulsar winds in
supernova remnants, radio afterglows of soft gamma-ray repeaters and
so forth.
\end{abstract}


\section{Introduction}

%
%

Many astrophysical processes, including star formation, stellar
collapse, supernova explosions, gamma-ray bursts (GRBs) and galaxy
cluster evolution etc., may involve self-gravitational inflows and
outflows during a certain phase of evolution. These problems with
different approximations (e.g., isothermal or polytropic equations
of state, spherical or cylindrical symmetries, radiative or
non-radiative regimes etc.) have been explored extensively in
various contexts (Bodenheimer \& Sweigart 1968; Shu 1977;
Goldreich \& Weber 1980; Suto \& Silk 1988; Foster \& Chevalier
1993; Boily \& Lynden-Bell 1995; Murakami, Nishihara \& Hanawa
2004; Shadmehri 2005).
When a gas medium has evolved
sufficiently far away from the influence of initial and boundary
conditions, it may gradually evolve into self-similar behaviours
(e.g., Sedov 1959; Landau \& Lifshitz 1959; Barenblatt \&
Zel'dovich 1972). This evolving trend as indicated by numerical
calculations to adjust to a roughly self-similar profile led
Larson (1969a,b) and Penston (1969a,b) to develop a spherically
symmetric isothermal self-gravitating similarity solution which
has a flat inner density profile with a radial speed proportional
to radius and a supersonic outer envelope with an expansion speed
3.3 times the isothermal sound speed. Shu (1977) constructed an
alternative self-similar expansion-wave collapse solution (EWCS),
in which the initial condition is an isothermal sphere with a gas
mass density $\rho\propto r^{-2}$ throughout in a hydrostatic
equilibrium. Perturbations or some central energy loss initiate
the collapse and a rarefacetion wave front travels outward through
the gas medium inside of which gas materials rapidly attain
free-fall velocity towards to the centre. This EWCS scenario of
inside-out collapse for protostar formation has been advocated by
\cite{sal87} and compared with observations
\citep{als,zhou93,choi95,sa99,harvey01}. Immediately following the
analysis of Shu (1977), Hunter (1977) constructed complete
isothermal self-similar solutions by going from negative times to
positive times and proposed a new matching method to find
solutions in density-speed phase diagrams. He found an infinite
number of discrete solutions in the `complete solution space'.
Whitworth \& Summers (1985) have further expanded these solutions
into two-parameter continua by allowing weak discontinuities
across the sonic critical line. Expansion wave solutions for more
general adiabatic and self-gravitating gas flows were discussed by
Cheng (1978). Suto \& Silk (1988) studied generalized polytropic
similarity solutions. Yahil (1983) obtained polytropic flow
solutions of the Larson-Penston (LP) type with the polytropic
index $\gamma$ in the range of $6/5<\gamma<4/3$ for applications
to stellar collapse and noted that various solution features
appear to be in accordance with results of numerical simulations.
\citet{fosterchevalier93} studied the gravitational collapse of an
isothermal cloud by hydrodynamic simulations. They recovered the
LP solution in the central region where a core forms and the
self-similar solution of \citet{shu77} when the ratio of initial
outer cloud radius to core radius is $\gsim 20$.

Shock processes can naturally occur in diverse astrophysical settings,
for example, supernova explosions, photoionized gas, stellar winds,
collisions between high velocity clumps of interstellar gas, collisions
of two or several galaxies etc. Shocks have been under extensive
investigations in contexts of supernova remnants (SNRs), nebulae
associated with T Tauri stars, planetary nebulae, HII regions,
particle accelerations and GRBs etc. (e.g., McKee \& Hollenbach 1980;
Draine \& McKee 1993; $\mathrm{M\acute{e}sz\acute{a}ros}$ 2002).
McKee \& Ostriker (1977) explored the interaction of a SNR with
an inhomogeneous ambient medium. Self-similar hydrodynamic shocks were
studied for interaction zones of colliding stellar winds (Chevalier 1982;
Chevalier \& Imamura 1983). Tsai \& Hsu (1995) constructed self-similar
shocks describing a situation in which a central thermal or kinetic
energy release initiates an outgoing shock during a protostellar
collapse to form a low-mass star. Tsai \& Hsu (1995) also constructed
a shock expansion solution with a finite core density matched with a
static SIS envelope; this solution has been generalized into a family
of `champagne flow' shock solutions by Shu et al. (2002) to model
expansions of HII regions surrounding massive OB stars after a rapid
passage of an initial ionization front in a neutral hydrogen cloud.
Shen \& Lou (2004) studied EECC shock solutions by allowing outflows
or inflows far away from the central region. In terms of modelling
protostellar systems, this flexibility can accommodate a variety of
physical possibilities. Bian \& Lou (2005) explored the parameter
space more systematically to construct various isothermal shock
solutions, including twin shocks, two temperature shocks and so forth.

Chevalier \& Imamura (1982) performed a linear stability analysis
for a one-dimensional shock driven by a constant velocity piston.
For a radiative cooling function $\Lambda\propto\rho^2T^{\alpha}$,
they found that radiative shocks were unstable for $\alpha\leq 0.4$
in a fundamental mode and unstable to overtone modes for
$\alpha\leqslant0.8$. Bertschinger (1986) further extended their
work to three dimensions and showed that transverse modes would be
unstable for $\alpha\leqslant1.0$. Ryu \& Vishniac (1987) applied
the linear theory to the dynamic instability of strong plane-parallel
or spherical adiabatic blast shock waves in a gas medium with an
initially uniform density and found that such blast waves to be
unstable for an adiabatic index $\gamma\leq1.2$. Blondin \& Cioffi
(1989) have shown that local instabilities are restricted to
wavelengths less than the shock thickness. Vishniac (1993) examined
the dynamical and gravitational instabilities of spherical shocks.
Toth \& Draine (1993) carried out linear stability analysis as well
as numerical simulations to study effects of a transverse magnetic
field on magnetohydrodynamic (MHD) shock stability and found that a
transverse magnetic field tends to stabilize the flow. MHD shocks in
a gas medium of low fractional ionization are subject to a novel
dynamical instability involving the deformation of magnetic field
lines (Wardle 1990). Decelerating radiative shock fronts (e.g.,
expanding SNRs in the snowplow phase) are subject to
a ripping instability (e.g., Vishniac 1983; Vishniac \& Ryu 1989).
Drury (1984) noted that acoustic waves propagating towards a shock
in the preshock medium may be amplified as they enter a region with
a steep cosmic-ray pressure gradient. Rayleigh-Taylor instabilities
arise in the cosmic-ray precursor would further complicate the
structure of the precursor when the density gradient due to these
nonlinear sound waves opposes the cosmic-ray pressure gradient
(e.g., Ryu 1993).

Magnetic fields play important roles in various astrophysical
environments. Complex filamentary structures in
molecular clouds, shapes and shaping of planetary nebulae,
synchrotron radiation from SNRs,
magnetized stellar winds, galactic winds, GRBs, dynamo effects in
stars, galaxies, and galaxy clusters as well as other interesting
problems all involve magnetic fields (e.g., Hartmann 1998; Balick
\& Frank 2002). Due to the formation of filamentary structures in
numerical simulations [e.g., \citet{porter94,klessen00,
ostriker01}] and the appearance of numerous filaments in
observations (e.g., Falgarone et al. 2001), analyses on processes
of filament formation and evolution have also been pursued
\citep{kawachi98,hennebelle03,tp03,shadmehri05}. For these
reasons, MHD shocks are under extensive investigations. Bazer \&
Ericson (1959) were among the first to study hydromagnetic shocks
for astrophysical applications. Whang (1984) studied
forward-reverse hydromagnetic shock pairs in the heliosphere.
Chakrabarti (1989, 1990) investigated MHD shocks in accretion
discs. Numerical simulations were carried out to investigate the
spherically symmetric shock interaction of pulsar wind nebula in a
SNR (e.g., van der Swaluw et al. 2001). Recently, Del Zanna et al.
(2004) re-examined this problem in an axisymmetric geometry. Ouyed
\& Pudritz (1993) studied oblique MHD shocks in disc winds from
young stellar objects (YSOs) to explain broad, blueshifted
forbidden emission lines observed in these sources.
Duncan \& Thompson (1992) proposed very strongly magnetized neutron
star might be the origin of soft SGRs and anomalous X-ray pulsars
(AXPs). Magnetic fields play an important role in the formation of
the three-ring structure around supernova 1987A (e.g., Tanaka \&
Washimi 2002). Gaensler et al. (2005) found an expanding radio
nebula produced by a giant flare from the magnetar SGR 1806-20 and
interpreted it as ejecta colliding with pre-existing shells.

In addition to Newtonian shocks, relativistic shocks are also
investigated in various contexts. Blandford \& McKee (1976) studied
relativistic blast wave solutions in spherical geometry and calculate the
active galactic nucleus (AGN) radiation spectrum using their blast wave
solutions. Kennel \& Coroniti (1984) considered ultra-relativistic MHD
shock models for the Crab Nebula. Emmering \& Chevalier (1987) studied
relativistic MHD shocks and applied their results to pulsar winds. Best
\& Sari (2000) re-examined the problem of Blandford \& McKee (1977) and
found second-type self-similar solutions to the problem of
ultra-relativistic strong explosions. Perna \& Vietri (2002) studied
the propagation of a relativistic shock in an exponential atmosphere
and applied their solutions to GRBs. Nakayama \&
Shigeyama (2005) reconsidered this problem in a plane-parallel geometry.
Hidalgo \& Mendoza (2005) considered imploding self-similar relativistic
shock waves. Dynamics of relativistic magnetized blast wave was explored
by Lyutikov (2002). Takahashi et al. (2002) investigated MHD adiabatic
shock accreting onto a rotating Kerr black hole. Subsequently, Fukumura
(2004) studied isothermal shock formation around a Kerr black hole. Das
et al. (2003) studied isothermal shocks around Schwarzschild black hole
using the pseudo-Schwarzschild potential. Mildly magnetized internal
relativistic shocks in GRBs have been invoked to explain the GRB prompt
emission data (Fan et al. 2004).

%
%
%
%
%

The role of a random magnetic field in our model analysis is
characterized by an important magnetic parameter $\lambda$ as
defined by equation (\ref{intro}). Different systems of
astrophysical objects involve a range of $\lambda$ values. The
estimated $\lambda$ parameter for the Crab Nebula is of the order
of a magnitude around $10^{5}\sim10^{6}$. For star forming
regions, the estimated $\lambda$ is typically in the range of
$0.01\sim0.1$. For planetary nebulae, a typical $\lambda$ is
approximately $10^{-3}$. For a cluster of galaxies, the estimated
$\lambda$ is approximately 0.2. We would take $\lambda\sim 0.1$
typical for a star formation region for our later discussions. For
a star formation cloud, we esitmate
\[
\lambda\sim 0.1
\bigg(\frac{B_{\parallel}}{1.34\times10^{-4}\mathrm{G}}\bigg)^2
\bigg(\frac{\rho}{5\times10^{-19}\mathrm{g/cm^3}}\bigg)^{-2}
\]
\[
\qquad\qquad
\times\bigg(\frac{r}{2.24\times10^{17}\mathrm{cm}}\bigg)^{-2}
\]

This paper is structured as follows. The basic nonlinear MHD
equations and the self-similar MHD transformation are described
in Section 2. In Section 3, we present the MHD shock conditions.
In Section 4, we solve the nonlinear MHD ordinary differential
equations (ODEs) numerically to construct various MHD shock
solutions. We provide comments and discussions in Section 5.
Relevant technical details of mathematical analyses are
summarized in several appendices for the convenience of
reference.

\section[]{PHYSICAL ASSUMPTIONS AND\\
\quad BASIC MHD MODEL FORMULATION}

For a random tangled magnetic field on small scales, we
formulate a large-scale MHD problem under the approximation
of quasi-spherical symmetry. In spherical polar coordinates
$(r,\ \theta,\ \phi)$, the mass conservation equation is
\begin{equation} \label{continuity}
\frac{\partial\rho}{\partial t}+\frac{1}{r^{2}}
\frac{\partial(r^{2}\rho u)}{\partial r}=0\ ,
\end{equation}
where $u$ is the radial bulk flow speed and $\rho$ is the gas
mass density. This mass continuity equation (\ref{continuity})
above is equivalent to the following equations, namely
\begin{equation}\label{muequivalent}
\frac{\partial M}{\partial r}=4\pi\rho r^{2}\ ,
\qquad\qquad
\frac{\partial M}{\partial t}
+u\frac{\partial M}{\partial r}=0\ ,
\end{equation}
where $M(r,t)$ is the enclosed mass within $r$ at time $t$.
The isothermal radial MHD momentum equation is simply
\begin{equation} \label{momentum}
\frac{\partial u}{\partial t}+u\frac{\partial u}{\partial r}
=-\frac{a^{2}}{\rho}\frac{\partial \rho }{\partial r}-
\frac{GM}{r^{2}}-\frac{1}{8\pi\rho}\frac{\partial}{\partial r}
<B^{2}_{\parallel}>-\frac{<B^{2}_{\parallel}>}{4\pi\rho r}
\end{equation}
%
%
where $a$ is the isothermal sound speed, ${\bf B_{\parallel}}=
(B^{2}_{\theta}+B^{2}_{\phi})^{1/2}$ stands for the random magnetic
field component parallel to the MHD shock front and $-\partial\Phi/
\partial r\equiv -GM(r,t)/r^2$ with $\Phi(r,t)$ being the gravitational
potential. Here, $<B^{2}_{\parallel}>$ represents a kind of ensemble
average of random magnetic field. In our formulation, some terms
involving cross correlations of radial and transverse magnetic field
components have been ignored (see Appendix C of Yu \& Lou 2005). Under
the quasi-spherical symmetry, the Poisson equation relating the gas
mass density $\rho$ and the gravitational potential $\Phi$ is
automatically satisfied. Here in equation (\ref{momentum}), we keep
the magnetic tension force term that was ignored in equation (3) of
\citet{chiueh}.

By taking the electrical conductivity to be infinite
(i.e., the ideal MHD approximation or complete frozen-in
approximation), we arrive at the following relation
\begin{equation}\label{rhoBr}
B_{\parallel}/(\rho r)=\textrm{const}\ ,
\end{equation}
where the right-hand side is a constant of integration. The
derivation of equation (\ref{rhoBr}) involves the mass
conservation and the transverse components of the magnetic
induction equation in the quasi-spherical approximation.
For details, we refer the reader to Yu \& Lou (2005).
Using integral (\ref{rhoBr}), we define the
dimensionless parameter $\lambda$ according to
\begin{equation}\label{intro}
\lambda\equiv B^{2}_{\parallel}/(16\pi^2 G\rho^{2}r^{2})
\end{equation}
which is a measure for the relative magnitudes of the magnetic
energy density and the self gravitational energy density.
In comparison with the previous work of Chiueh \& Chou (1994),
it is apparent that
\begin{equation}\label{beta}
\lambda={\beta(x)}/({\alpha^{2}x^{2}})\ ,
\end{equation}
where $\alpha(x)$ and $x$ are the reduced density and the
independent similarity variable defined immediately below
in equation (\ref{reducedvarialbe}), respectively. Here,
variable $\beta(x)$ is a dimensionless function introduced
by \citet{chiueh} and stands for the reduced magnetic
pressure, or equivalently, the reduced magnetic energy
density. With this comparison, we know from their equation
(11) that while our formulation differs from theirs in
several aspects, the final magnetosonic critical conditions
are in fact identical. The ratio of the Alfv\'en wave speed
$v_A$ to the isothermal sound speed $a$ is given by
\[
\qquad \frac{v_{A}}{a}=\bigg(\frac{\beta}{\alpha}\bigg)^{1/2}
\qquad\hbox{ with }\qquad
v_{A}\equiv\frac{B_{\parallel}}{(4\pi\rho)^{1/2}}\ .
\]
%
%
%
The dimensionless independent similarity variable is
defined by $x\equiv r/(at)$ and the self-similar MHD
transformation are
\[
\rho(r,t)=\frac{\alpha(x)}{4\pi G t^{2}}\ , \quad
M(r,t)=\frac{a^{3}t}{G}m(x)\ , \quad u(r,t)=a v(x)\ ,
\]
\begin{equation}\label{reducedvarialbe}
\qquad \Phi(r,t)=a^{2}\phi(x),
\qquad\qquad
B_{\parallel}(r,t)=\frac{a b(x)}{\sqrt{G}t}\ ,
\end{equation}
where the dimensionless $\alpha(x)$, $m(x)$, $v(x)$, $\phi(x)$ and
$b(x)$ are the reduced dependent variables for the mass density,
the enclosed mass, the radial flow speed, the gravitational
potential and the transverse magnetic field strength, respectively;
and they are all dimensionless functions of $x$ only. Note that
$b(x)=\sqrt{\lambda}\alpha x$ and is related to the dimensionless
function $\beta(x)$ of \citet{chiueh} by $b^{2}\equiv\beta$.

It follows that equations (\ref{muequivalent}) reduce to
\begin{equation}\label{muequiv1}
m^{'}=x^{2}\alpha\ , \qquad\qquad (v-x)m^{'}+m=0\ ,
\end{equation}
where the prime denotes the derivative with respect
to $x$. Combining these two ODEs, we immediately
obtain the following two equations
\begin{equation}\label{mpositive}
m=(x-v)x^{2}\alpha\ , \qquad\qquad
[x^{2}\alpha(x-v)]^{'}=x^{2}\alpha\ .
\end{equation}
By equation (\ref{mpositive}), the physical requirement
of a positive enclosed mass $m(x)>0$ is equivalent to
the condition $x-v>0$. Thus, solutions of $v$ must lie
to the upper-right of the straight line $x-v=0$ in the
plane $-v(x)$ versus $x$.

After substituting the self-similar variables into continuity
and momentum equations, we derive the following two coupled
nonlinear MHD ODEs
\[
\big[(x-v)^{2}-(1+\lambda\alpha x^{2})\big]v^{'}
\]
\begin{equation}\label{mainv}
\qquad\qquad\qquad\qquad=(x-v)\big[\alpha(x-v)-2/x\big]\ ,
\end{equation}
\[
\big[(x-v)^{2}-(1+\lambda\alpha x^{2})\big]\alpha^{'}/\alpha
\]
\begin{equation}\label{mainalpha}
\qquad\qquad\qquad=(x-v)\big[\alpha-2(x-v)/x\big]
+2\lambda x\alpha\ .
\end{equation}
Setting $\lambda=0$ for the absence of the magnetic field, we
readily recover the hydrodynamic formulation of Lou \& Shen (2004).
In the above two coupled nonlinear MHD ODEs (\ref{mainv}) and
(\ref{mainalpha}), the magnetosonic critical curve is specified by
\begin{equation}\label{critical}
x-v=2/(\alpha x)
\qquad\hbox{ and }\qquad
x-v+2\lambda x=(x-v)^{3}\ .
\end{equation}
Note that for $\lambda=0$ and $x-v>0$, equation (\ref{critical})
becomes $x-v=1$ and $\alpha=2/x$. Given the positiveness of $x-v>0$, we
come to the familiar condition $x-v=1$ for the isothermal sonic critical
line \citep{loushen}. Equation $(x-v)^{2}=1+\lambda\alpha x^{2}$ is
equivalent to the equation $(x-v)^{2}=(a^{2}+v^{2}_{A})/a^{2}$,
directly related to the fast magnetosonic condition as anticipated
on the ground of physics.

The asymptotic MHD solution behaviours of the coupled nonlinear
MHD ODEs are summarized below. As MHD generalizations, these
asymptotic solutions were derived in parallel to hydrodynamic
results (see Appendix A of Lou \& Shen 2004 for more details of
their derivations). We emphasize that there is no hydrodynamic
counterpart for class III asymptotic solutions as
$x\rightarrow 0^{+}$ described below.
In addition to $v(x)$ and $\alpha(x)$, we also provide
corresponding asymptotic solutions of $\beta(x)$ for
the reduced magnetic pressure for relevant physical
information and a more complete presentation.

In the limit of $x\rightarrow+\infty$,
we have asymptotic solutions
\[
v(x)=V+\frac{2-A}{x}+\frac{V}{x^{2}}
\]
\begin{equation}\label{shockinftyv}
\quad+\frac{(A/6-1)(A-2)+2V^{2}/3
+A(2-A)\lambda/3}{x^{3}}+\cdots\ , \label{vinfty}
\end{equation}
\[
\alpha(x)=\frac{A}{x^{2}}
+\frac{A(2-A)}{2x^{4}}+\frac{A(4-A)V}{3x^{5}}+
\]
\[
\frac{A(A-3)(A/2-1)-(A-6)A V^{2}/4+(2-A)A^{2}\lambda/4}{x^{6}}
\]
\begin{equation}\label{shockinftya}
\hspace{2em}+\cdots\ ,
\end{equation}
\[
\beta(x)=\lambda\alpha^{2} x^{2}
=\frac{A^2\lambda}{x^2}+\frac{A^2
(2-A)\lambda}{x^4}+\frac{A^2(2-A)^2\lambda}{4x^6}
\]
\begin{equation}\label{shockinftyb}
\hspace{2em}+\cdots\ ,
\end{equation}
where $A$ and $V$ are two constants of integration
mainly for mass density and radial flow speed,
respectively.

Class I: For a central MHD free-fall collapse in the limit of
$x\rightarrow 0^{+}$, we have to the leading order
\begin{equation}\label{shockclassIv}
v= -2F/x^{1/2}-\frac{3}{4F}x^{1/2}\ln x-2L
x^{1/2}+\cdots \ ,
\end{equation}
\begin{equation}\label{shockclassIa}
\alpha= F/x^{3/2}-\frac{3}{8F}x^{-1/2}\ln x-L
x^{-1/2}+\cdots \ ,
\end{equation}
\begin{equation}
\beta=\lambda\alpha^{2} x^{2}=\frac{\lambda F^2}{x}+\cdots \ ,
\end{equation}
where $F$ and $L$ are two constants of integration.

Class II: For the central LP-type MHD solutions being regular as
$x\rightarrow 0^{+}$, we have
\begin{equation}\label{shockclassIIv}
v=\frac{2}{3}x+\frac{(2-3D-18D\lambda)}{135}x^{3}+\cdots\ ,
\end{equation}
\begin{equation}\label{shockclassIIa}
\alpha=D+\frac{D(2-3D-18D\lambda)}{18}x^{2}+\cdots\ ,
\end{equation}
\begin{equation}\label{shockclassIIb}
\beta=\lambda\alpha^2 x^2=\lambda D^2 x^2+\cdots\ ,
\end{equation}
where $D$ is an integration constant.

Class III: There exists a novel class of asymptotic MHD solutions
which requires a sufficiently strong magnetic field and would
disappear in the absence of magnetic field. As $x\rightarrow
0^{+}$, we write to the leading order
\begin{equation}\label{shockclassIIIv}
v(x)= H x+\cdots \ ,
\end{equation}
\begin{equation}\label{shockclassIIIa}
\alpha(x)= K/x^{\eta}+\cdots \ ,
\end{equation}
where $K$ is an arbitrary integration constant and
\begin{equation}\label{strongB}
H=\big[2-\lambda\pm(\lambda^2-4\lambda)^{1/2}\big]/2<0\ ,
\end{equation}
\begin{equation}
\eta=(1-H+2\lambda)/\lambda
\qquad\hbox{ with }\qquad 2\leq\eta\leq3\ .
\end{equation}
It is then straightforward to infer
that as $x\rightarrow 0^{+}$,
\begin{equation}
\beta(x)=\lambda\alpha^2 x^2
\rightarrow\frac{\lambda K^2}{x^{2(\eta-1)}}\ ,
\end{equation}
\begin{equation}
B_{\parallel}^2\rightarrow\frac{\lambda K^2 a^{2\eta}
t^{2(\eta-2)}}{G r^{2(\eta-1)}}\ .
\end{equation}
For this new class of asymptotic MHD solutions, the magnetic field
strength scales as $B_{\parallel}\propto r^{-(\eta-1)}$ at a fixed
moment $t$, while at a given radius $r$, we have the scaling
$B_{\parallel}\propto t^{\eta-2}$. The power index $\eta$ is such
that $-2\leq-(\eta-1)\leq-1$. By expression (\ref{strongB}), this
type of MHD solutions exists only when the magnetic parameter
$\lambda$ is greater than $4$.

It is possible for similarity MHD solutions to go across the
magnetosonic critical curve smoothly. Along the magnetosonic
critical curve (e.g., Jordan \& Smith 1977), the two MHD
eigensolutions are governed by the following quadratic
equation with $z\equiv v'$, namely
\begin{equation}\label{eigenderivative}
\bigg[(v-x)-\frac{x\lambda}{(v-x)^2}\bigg]z^2
+(x-v)z-\frac{v}{x^2}=0\ .
\end{equation}
From equation (\ref{eigenderivative}), we obtain two types of MHD
eigensolutions for $z\equiv v'$ along the magnetosonic critical curve.
When the two roots of equation (\ref{eigenderivative}) are of opposite
signs, type 1 and type 2 eigensolutions are those with negative and
positive roots of equation (\ref{eigenderivative}), respectively.
When the two roots of equation (\ref{eigenderivative}) are of the
same sign, type 1 and 2 eigensolutions are those with smaller and
larger absolute values respectively. In the open interval $0<x<2$,
type 1 and type 2 are exactly defined as such. When $x>2$ (relevant
to the LP-type MHD solution), $dv/dx$ of type 1 has a larger absolute
value, while $dv/dx$ of type 2 has a smaller absolute value; that
is, their magnitudes reverse for nodal points.
In our current definition, no magnitude reversal would happen. When
the point is a nodal point, type 1 and type 2 solutions remain always
the smaller and larger ones respectively for the absolute value of
$dv/dx$. To summarize, our definition is not defined by the explicit
expressions of $dv/dx$ (e.g., $1-1/x_{\ast}$ and $1/x_{\ast}$), but
by their magnitudes and signs. It is easier to keep in mind their
relevant physical properties.

\section[]{ISOTHERMAL MHD SHOCK CONDITIONS}

\subsection[]{Perpendicular MHD Shock}

\subsubsection[]{The One-Temperature Case}


The simplest type of MHD shock wave is the perpendicular shock. In this
case, the velocities of both the shock and plasma are perpendicular to
the magnetic field, which itself is parallel to the shock front. In a
frame of reference moving with the MHD shock front, the properties on
both sides of the shock $(\rho_1,u_1,B_1,p_1)$ and $(\rho_2,u_2,B_2,p_2)$
are related by the equations for conservation of mass, momentum, energy
and magnetic flux. In the isothermal approximation, we need not to
consider the MHD energy equation.

In dimensional form, the perpendicular MHD
shock conditions are (e.g., Priest 1982)
\begin{equation}
\rho_{2}u_{2}=\rho_{1}u_{1}\ ,
\end{equation}

\begin{equation}\label{onetempmomentum}
p_{2}+B_{2}^{2}/(8\pi)+\rho_{2}u_{2}^{2}
=p_{1}+B_{1}^{2}/(8\pi)+\rho_{1}u_{1}^{2}\ ,
\end{equation}
%
%

\begin{equation}\label{onetempmagflux}
B_{2}u_{2}=B_{1}u_{1}\ .
\end{equation}
In equation (\ref{onetempmomentum}), the term $B^2/(8\pi)$
represents the magnetic pressure. In equation
(\ref{onetempmagflux}), the quantity $Bv$ gives the rate at which
the magnetic flux is transported across a unit surface area;
physically, this quantity is actually proportional to the electric
field component tangential to the shock front and thus should
remain continuous across a shock. After straightforward
manipulations, the solution to the above set of MHD shock
equations can be expressed in terms of the mass density ratio $
\rho_{2}/\rho_{1}\equiv X,
$
the upstream Mach number
$
M_{1}\equiv u_{1}/a,
$
and the upstream plasma beta parameter\footnote{We emphasize
here that $\beta_1$ is the upstream plasma parameter, not the
dimensionless $\beta(x)$ function for the reduced magnetic
pressure defined in the previous context (see equation
\ref{beta}).}
$\beta_{1}\equiv p_{1}/(B_{1}^{2}/8\pi ) =2a^{2}/v_{A1}^{2}. $
The results are simply
\[
u_{2}/u_{1}=1/X\ ,
\]
\[
B_{2}/B_{1}=X\ ,
\]
\[
p_{2}/p_{1}=X\ ,
\]
where $X$ is the positive root of the following quadratic
equation (see Appendix A for detailed derivations)
\begin{equation}\label{onetempf}
f(X)=X^{2}+(\beta_{1}+1)X-\beta_{1}M_{1}^{2}=0\ ;
\end{equation}
this equation readily indicates the existence of only
one positive root of $X$. Physically, we should
further require $X>1$ for a fast magnetosonic shock.


When upstream conditions are specified, we can determine
the downstream conditions behind a magnetosonic shock
systematically. We simply take $\rho_{2}$ as $\rho_{d}$
and $\rho_{1}$ as $\rho_{u}$ such that
\[
u_{2}=u_{d}-u_{s}\ ,\qquad\qquad u_{1}=u_{u}-u_{s}\ ,
\]
\[
u_{u}=a v_{u}\ ,
\qquad\qquad u_{d}= a v_{d}\ ,
\qquad\qquad u_{s}= a x_{s}\ ,
\]
\[
X=\rho_{2}/\rho_{1}=\rho_{d}/\rho_{u}\ ,
\]
where $u_s$ is the radial shock speed, and subscripts $u$ and $d$
denote physical variables associated with upstream and downstream
sides, respectively. It then follows that
\[
\alpha_{d}/\alpha_{u}=X\ ,
\]
\[
(v_{d}-x_{s})/(v_{u}-x_{s})=1/X\ ,
\]
\[
M_{1}=v_{u}-x_{s}\ ,
\]
\[
\beta_{1}=2/(\lambda x_{s}^{2} \alpha_{u})\ .
\]
Here, the value of $X$ should be larger than $1$ for
a magnetosonic shock -- a kind of MHD fast shock.

The analysis of quadratic equation (\ref{onetempf})
shows that the two real roots of $X$ have opposite signs.
We naturally pick out the positive root among the two.

When $f(X=1)<0$, i.e., $2+\beta_{1}<\beta_{1}M_{1}^{2}$ or
$1+2/\beta_{1}<M_{1}^{2}$, it has a positive root $X>1$.
This conclusion can also be reached by directly solving the
quadratic equation and by requiring the positive root $X>1$.
Physically for a magnetosonic shock to exist, the upstream
flow speed relative to the shock speed must exceed the
upstream fast magnetosonic speed $(a^2+v_{A 1}^2)^{1/2}$.
This serves as a guide in constructing magnetosonic shocks.

When $f(X=1)>0$, i.e., $2+\beta_{1}>\beta_{1}M_{1}^{2}$ or
$1+2/\beta_{1}>M_{1}^{2}$, there is a positive root $X<1$.

Using the definitions of $\beta_{1}$ and $M_{1}$ in the
condition $1+2/\beta_{1}=M_{1}^{2}$ and dropping the subscript
$1$, this equation describes exactly the behaviour of the
magnetosonic critical curve $(x-v)^2=1+\lambda\alpha x^{2}$.





\subsubsection[]{Two-Temperature MHD Shocks}

When the constant temperatures on the two sides of
a magnetosonic shock are different, we have the
following two MHD jump conditions in the shock
framework of reference
\begin{equation}
\alpha_{d}(v_{d}-x_{sd})a_{d}=\alpha_{u}(v_{u}-x_{su})a_{u}\ ,
\end{equation}
\[
a_{d}^{2}\big[\lambda\alpha_{d}^{2}x_{sd}^{2}/2
+\alpha_{d}+\alpha_{d}(v_{d}-x_{sd})^{2}\big]
\]
\begin{equation}
\qquad\qquad=a_{u}^{2}\big[\lambda\alpha_{u}^{2}x_{su}^{2}/2
+\alpha_{u}+\alpha_{u}(v_{u}-x_{su})^{2}\big]\ .
\end{equation}
%
%
When the upstream conditions are given and we are going to
determine the downstream conditions, we take $\rho_{2}$ as
$\rho_{d}$ and $\rho_{1}$ as $\rho_{u}$. It then follows that
\begin{equation}
{a_{d}}/{a_{u}}=\tau\ , \qquad\qquad
{x_{sd}}/{x_{su}}={1}/{\tau}\ ,
\end{equation}
\begin{equation}
{\alpha_{d}}/{\alpha_{u}}
={(v_{u}-x_{su})}/{[\tau(v_{d}-x_{sd})]}=X\ ,
\end{equation}
%
\begin{equation}
\beta_{1}={2}/{(\lambda x_{su}^{2}\alpha_{u})}\ ,
\end{equation}
\begin{equation}\label{twoT}
{X^{2}}/{\beta_{1}}+\tau^{2}X+{M_{1}^{2}}/{X}=
{1}/{\beta_{1}}+1+M_{1}^{2}\ .
\end{equation}
Equation (\ref{twoT}) is a cubic equation in terms of $X$, namely
\begin{equation}\label{cubicXtau}
f(X)\equiv X^{3}+\beta_{1}\tau^{2}X^{2}
-(1+\beta_{1}+\beta_{1}M_{1}^{2})X+\beta_{1}M_{1}^{2}=0\ .
\end{equation}
For a physical magnetosonic shock solution, we should require
both $\tau>1$ and $X>1$. In constructing magnetosonic shock
solutions, we then choose the positive root(s) $X>1$ among
the three roots of cubic equation (\ref{twoT}) for $X$. Since
$f(X=1)=\beta_{1}(\tau^2-1)>0$
and $f(X=0)=\beta_{1}M_{1}^{2}>0$, cubic equation (\ref{twoT}) must
have at least one negative $X$ root because $f(-\infty)=-\infty$ and
$f(+\infty)=+\infty$. The remaining two roots could be either both
complex or both real. When the other two roots are either complex
roots or both less than 1, there is no magnetosonic shock solution.
When these other two real $X$ roots are both larger than 1, there
exist two possible magnetosonic shock solutions. From equation
(\ref{cubicXtau}), we derive $f'(X)\equiv df/dX$ as
\begin{equation}\label{primef}
f'(X)=3X^{2}+2\beta_{1}\tau^{2}X-(1+\beta_{1}+\beta_{1}M_{1}^{2})\ .
\end{equation}
According to expression (\ref{primef}), the two
real roots (one positive $X_{+}$ and the other
negative $X_{-}$) of $f'(X)=0$ are given by
\begin{equation}\label{Xpm}
X_{\pm}=\{-\beta_1\tau^2\pm [\beta_1^2\tau^4
+3(1+\beta_1+\beta_1M_1^2)]^{1/2}\}/3\ .
\end{equation}
By definition (\ref{cubicXtau}) for $f(X)$ and the fact of
$f(X=0)=\beta_{1}M_{1}^{2}>0$, it is clear that $f(X_{-})$
should remain positive. For the situation of $f(X_{+})>0$,
the two remaining roots form a complex conjugate pair and
there is no magnetosonic shock solution. For the situation
of $f(X_{+})\leq 0$, the two remaining roots are real and positive.
It is found that if the two roots are real only two situations
can happen: (i) the two roots are both positive and less than 1;
and (ii) the two roots are both larger than 1. The situation
that one root is greater than 1 and the other positive root is
less than 1 cannot occur because this would demand $f(X=1)<0$.


We have just analyzed several solution properties of the cubic
equation for the mass density ratio $X$ given upstream physical
conditions of a magnetosonic shock. However, in practical
calculations, we did not search for root $X>1$ using the above
procedure with $\tau>1$. We adopted an alternative yet
equivalent procedure described below.
In fact, our later two-temperature calculations are
performed with the following procedure involving one
root of an equivalent cubic equation.

Reciprocally, when downstream conditions are given, we can calculate
the upstream conditions across a magnetosonic shock. Under this
situation, we simply take $\rho_{2}$ as $\rho_{u}$ and $\rho_{1}$
as $\rho_{d}$ and so forth. It then follows that
\begin{equation} \frac{a_{u}}{a_{d}}=\tilde\tau\ ,
\qquad\qquad
\frac{x_{su}}{x_{sd}}=\frac{1}{\tilde\tau}\ ,
\end{equation}
\begin{equation}
\frac{\alpha_{u}}{\alpha_{d}}
=\frac{(v_{d}-x_{sd})}{\tilde\tau(v_{u}-x_{su})}=\tilde X\ ,
\end{equation}
\[
\tilde M_1=v_d-x_{sd}\ ,
\]
\[
\tilde\beta_{1}=2/(\lambda x_{sd}^{2} \alpha_{d})\ ,
\]
\begin{equation}
{\tilde X^{2}}/{\tilde\beta_{1}}
+\tilde\tau^{2}\tilde X+{\tilde M_{1}^{2}}/{\tilde X}=
{1}/{\tilde\beta_{1}}+1+\tilde M_{1}^{2}\ .
\end{equation}
This is a cubic equation in $\tilde X$ of exactly the same
mathematical form as equation (\ref{cubicXtau}), namely
\begin{equation}\label{tilde}
f(\tilde X)=\tilde X^{3}+\tilde\beta_{1}\tilde\tau^{2}
\tilde X^{2}-(1+\tilde\beta_{1}+\tilde\beta_{1}\tilde M_{1}^{2})
\tilde X+\tilde\beta_{1}\tilde M_{1}^{2}=0\ .
\end{equation}
For a physical magnetosonic shock, we naturally require $\tilde\tau<1$
and $\tilde X<1$. In numerical calculations, we choose the positive
$\tilde X$ root that is less than 1 among the three roots of cubic
equation (\ref{tilde}). It turns out to be much simpler in analyzing
this cubic equation. Because here
$f(\tilde X=1)=\tilde\beta_{1}(\tilde\tau^2-1)<0$ and
$f(\tilde X=0)=\tilde\beta_{1}\tilde M_{1}^{2}>0$, this cubic equation
must then have three real roots, one negative and two positive. One
positive root is larger than 1 and the other positive root is smaller
than 1. In comparison with the cases of $\tau>1$ in equation
(\ref{cubicXtau}), there is only one $\tilde X$ root of cubic equation
(\ref{tilde}) physically corresponding to a magnetosonic shock solution.
That the roots of the $\tau>1$ cases in equation (\ref{cubicXtau}) are
more complex is related to the fact that $f(X=1)=\beta_{1}(\tau^2-1)>0$.
Fortunately, in our specific model calculations, we just compute the
$\tilde\tau<1$ cases and the amount of our numerical computations can
be greatly reduced.

We have also considered the case of oblique MHD shocks in more details
(not completely shown here) and find that oblique MHD shocks do not
exist in our current model framework. The main reason that we cannot
have oblique MHD shocks in our formalism is related to the frozen-in
condition (\ref{rhoBr}).
Thus from the two following shock relations
for an oblique MHD shock (Priest 1982)
\[
\frac{B_{2y}}{B_{1y}}=\frac{(u_1^2-v_{A1}^2)X}
{(u_1^2-X v_{A1}^2)}\
\]
and
\[
X=\frac{\rho_{2}}{\rho_{1}}\ ,
\]
together with $B_{\parallel}\propto\rho$,
we come to the conclusion that
\[
\frac{(v_1^2-v_{A1}^2)X}{(v_1^2-X v_{A1}^2)}=X \ .
\]
There is only a trivial solution with $X=1$. From now on, we
ignore the radial magnetic field component and just consider
perpendicular MHD shocks.

\section[]{SELF-SIMILAR ISOTHERMAL\\
\quad\  MHD SHOCK SOLUTIONS}

In this section, we present and discuss properties
of semi-complete global solutions with perpendicular
MHD shocks.

\subsection[]{Extensions of Previous Shock
Results by Including a Random Magnetic Field}

With the inclusion of a random magnetic field, we can
extend previous hydrodynamic similarity shock solutions.
More specifically, we would extend the results of Tsai
\& Hsu (1995), Shu et al. (2002) and Shen \& Lou (2004).
For a random magnetic field, we envision a simple ``ball of
thread" scenario in a huge spatial volume of gas medium. A
magnetic field line follows the `thread' meandering within
a thin spherical `layer' in space in a random manner. In the
strict sense, there is always a random weak radial magnetic
field component such that random magnetic field lines in
adjacent `layers' are actually connected throughout in space.
By taking a large-scale ensemble average of such a magnetized
system, we are then left with `layers' of random magnetic
field components transverse to the radial direction. Very
recently, Bian \& Lou (2005) carried out extensive
investigations on self-similar isothermal shock solutions
in the absence of magnetic field, while under the
approximation of a quasi-spherical symmetry in the sense
described above, Yu \& Lou (2005) constructed smooth
self-similar isothermal MHD solutions involving a random
magnetic field but without shocks. To a greater extent, our
present investigation is to combine the analyses of Bian \&
Lou (2005) and Yu \& Lou (2005) as well as to explore new
possibilities associated with magnetic field.

\subsubsection[]{MHD Extensions for Shocks of Tsai \& Hsu}

\begin{figure}
 \includegraphics[width=3in]{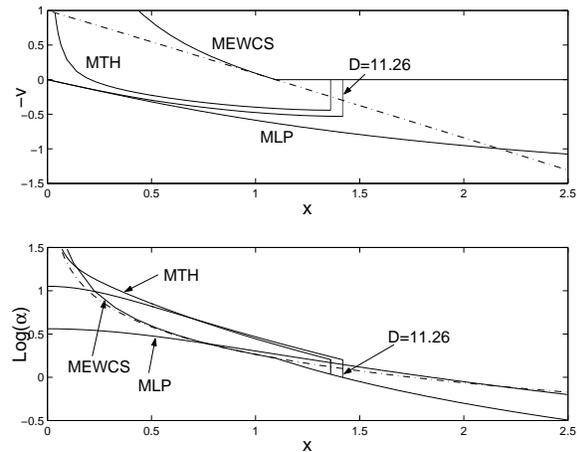}
 \caption{\label{tsaihsu} The MHD case of $\lambda=0.1$. The MHD
 extensions of Tsai \& Hsu (MTH) type similarity shock solutions
 (Class I MHD shock with free-fall diverging behaviour near the
 origin and Class II MHD shock with a finite reduced density near
 the origin and an eigenvalue $D=11.26$). The negative reduced
 radial velocity $-v(x)$ (upper panel) and the reduced density
 $\alpha(x)$ in logarithmic scale (lower panel) versus $x$ in linear
 scale are shown respectively. MHD Expansion-Wave Collapse Solution
 (MEWCS) as well as MHD LP (MLP) solution are also displayed.
 Dash-dotted line represents the magnetosonic critical curve.
 Class I MHD Tsai \& Hsu (MTH) type similarity shock location is at
 $x_{s}=1.3599$ and this solution passes through the magnetosonic
 critical curve at $x_{*}=0.0369$. Class II MHD Tsai \& Hsu (MTH)
 type similarity shock location is at $x_{s}=1.4194$.
  }
\end{figure}

\begin{figure}
 \includegraphics[width=3in]{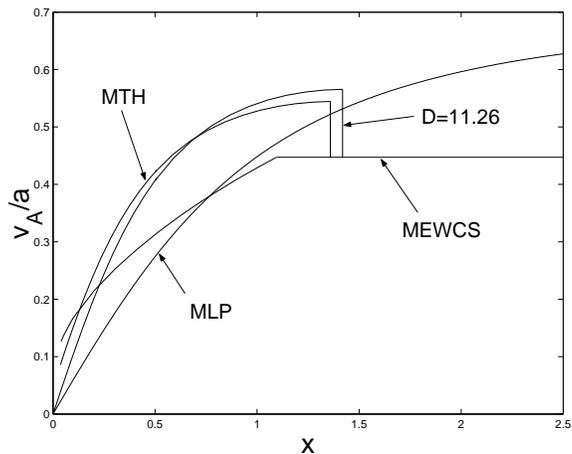}
 \caption{\label{alfventsaihsu} The MHD case of $\lambda=0.1$. The
 MHD extensions of Tsai \& Hsu (MTH) type similarity shock solutions.
 The ratio of Alfv\'en wave speed to the isothermal sound speed $v_A/a$
 versus $x$ is shown in linear scales. MHD Expansion-Wave Collapse
 Solutions (MEWCSs) and MHD LP (MLP) shock solutions are displayed.
  }
\end{figure}

Alternative to the results of Larson (1969a, b) and Penston (1969a, b),
Shu (1977) constructed the EWCS and developed an inside-out collapse
scenario for the process of forming low-mass protostars (e.g., Shu et
al. 1987). Tsai \& Hsu (1995) considered a self-similar shock
travelling into a static SIS envelope characterized by $\tau=1$,
$x_{sd}=x_{su}=x_{s}$, $v_{u}=0$ and $\alpha_{u}=2/x_{s}^{2}$. Their
shock connection condition becomes $v_{d}=x_{s}-1/x_{s}$ and
$\alpha_{d}=2$ (see their equation 16). Their global shock solutions
have different asymptotic behaviours near the origin (see their fig. 6).
The diverging free-fall solution near the origin is the Class I solution.
The converging LP-type solution near the origin is the Class II solution.
In our formalism, we can generalize the shock results of Tsai \& Hsu
(1995) by incorporating a random magnetic field. When a magnetic field is
included, the MHD Class I shock solution generalizing that of Tsai \& Hsu
(1995) crosses the magnetosonic critical curve at $x_{*}=0.0369$ with the
shock location at $x_{s}=1.3599$ for an outgoing MHD shock front travelling
at a constant speed of 1.3599 times the isothermal sound speed $a$. The
corresponding central mass accretion rate is 0.0726.
For the MHD Class II shock solution, a central expansion with a finite
central density (LP-type solution) matches with a magnetostatic SIS
envelope across an MHD shock. The shock is located at $x_{s}=1.4194$
and thus shows a higher shock speed of $1.4194$ times the isothermal
sound speed $a$ and a reduced core density $D=11.26$ [see asymptotic
solutions (\ref{shockclassIIv}), (\ref{shockclassIIa}), and
(\ref{shockclassIIb})]. In Figure \ref{tsaihsu} and Figure
\ref{alfventsaihsu}, we plot the MHD extensions of Tsai \& Hsu type
self-similar shocks. Note that these MHD generalizations of Tsai \&
Hsu shock solutions are just two special cases among magnetized
similarity shock solutions into a magnetostatic SIS envelope. Details
of the procedure for constructing these MHD shock solutions will be
discussed in section 4.2.
%

\subsubsection[]{MHD Extensions for Shock Solutions of Shu et al.}

\begin{figure}
\begin{center}
\includegraphics[scale=0.3]{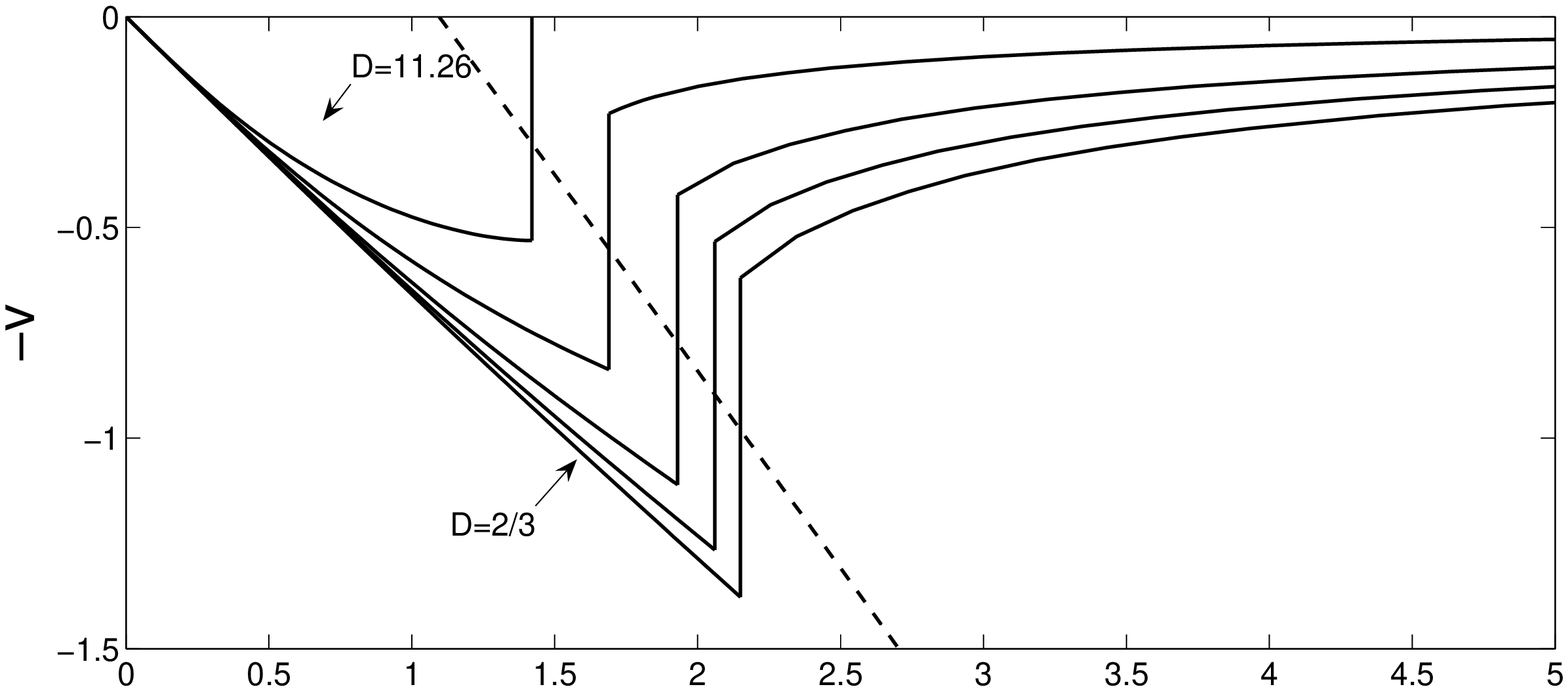}%
\hspace{1in}%
\includegraphics[scale=0.3]{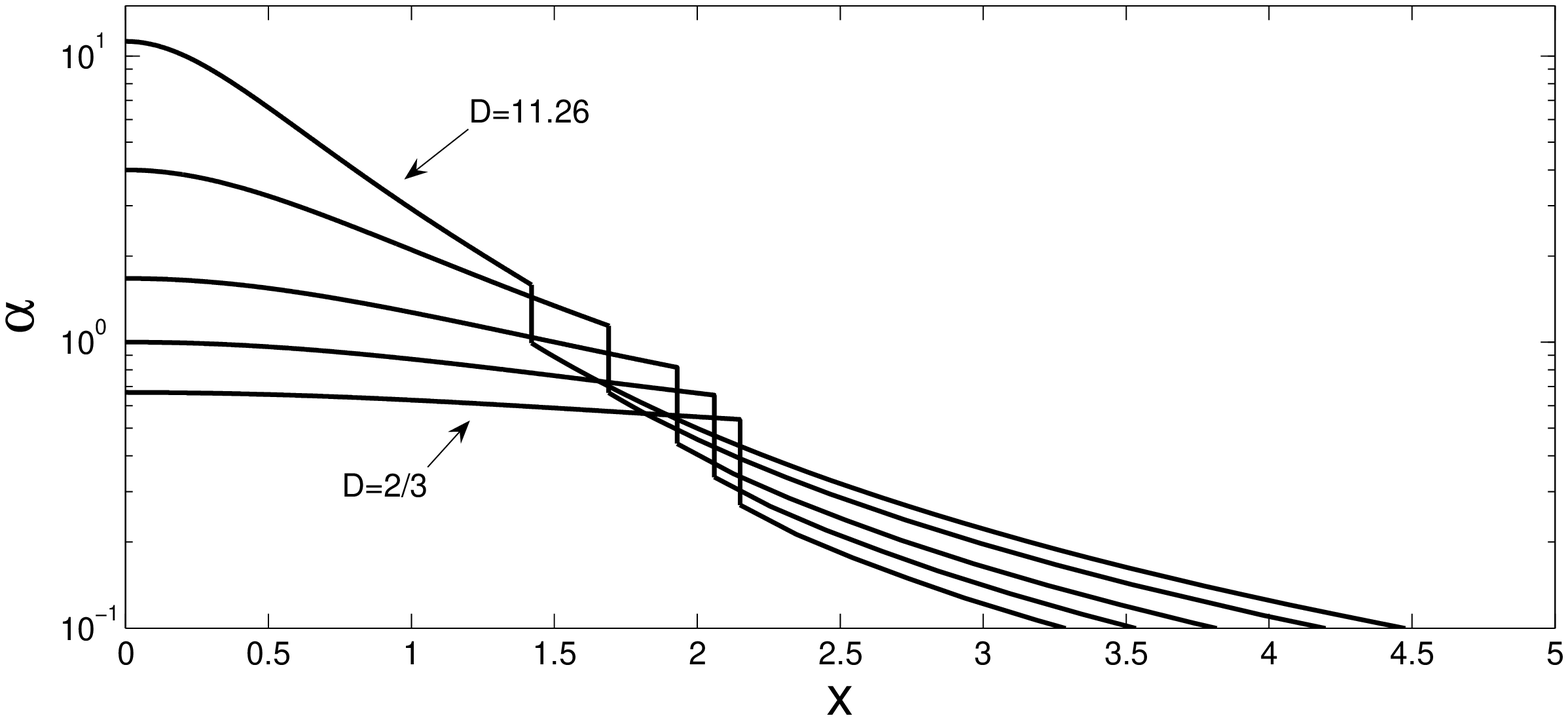}
\caption{\label{shutype}
 The MHD generalization for similarity shock solutions of Shu et al.
 (2002) with $\lambda=0.1$. For self-similar MHD shock solution
 curves from bottom to top, $D=2/3,\ 1,\ 1.67,\ 4,\ 11.26$,
 respectively. The negative reduced radial speed $-v(x)$ in linear
 scale (upper panel) and the reduced mass density $\alpha(x)$ in
 logarithmic scale (lower panel) versus $x$ in linear scale. Both
 the MHD shock speed and strength increase with a decreasing $D$
 parameter. As $D$ approaches $0^{+}$, the reduced speed $v(x)$
 converges to an invariant form with an invariant fastest and
 strongest shock located at $x_{s}=2.56$ (see also the heavy
 solid curve in Fig. \ref{champagne}). Dashed line represents
 the magnetosonic critical curve.}
\end{center}
\end{figure}

\begin{figure}
\includegraphics[scale=0.4]{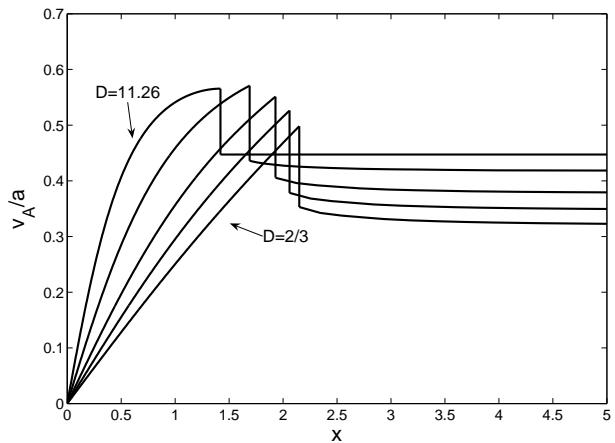}
 \caption{\label{alfvenshutype}
 MHD extensions for similarity shock solutions of Shu et
 al. (2002) with $\lambda=0.1$. The ratio of Alfv\'en
 wave speed and isothermal sound speed $v_A/a$ versus $x$
 is shown. Both the shock speed and strength increase with
 a decreasing $D$ parameter.  }
\end{figure}

Shu et al. (2002) extended the Class II shock solution
of Tsai \& Hsu (1995) and used these solutions to model
`champagne flows' of HII regions surrounding massive OB stars.
In a similar manner, by varying the $D$ and $x_s$ parameters with
$V=0$ and a fixed $\lambda$, MHD extensions for shock solutions of
Shu et al. (2002) are also constructed and presented in Figure
\ref{shutype} and Figure \ref{alfvenshutype}. Here the parameters
$D$, $V$, $\lambda$ and $A$, $K$ and $H$ in later discussion are
all integration constants in asymptotic solutions (\ref{shockinftyv})
to (\ref{shockclassIIIa}). Both the shock speed and shock strength
increase with a decreasing $D$ parameter. In the limit of
$D\rightarrow 0^{+}$ numerically, the mass parameter $A$ approaches
$0^{+}$ accordingly and the reduced radial speed $v(x)$ converges
to an invariant form (Shu et al. 2002; Shen \& Lou 2004) with an
invariant fastest and strongest MHD shock located at
$x_s=2.56$. In other words,
as $D\rightarrow 0^{+}$ numerically, behaviours of the MHD shock
solutions gradually become the same as those of the hydrodynamic
shock solutions, and for the same parameter $D$, the MHD shock
speeds are faster than the hydrodynamic shock speeds.
These shock solutions are MHD generalizations of `champagne
breezes' (see curves in our Fig. \ref{shutype} and the heavy
solid curve in our Fig. \ref{champagne} and compare with fig. 1
of Shu et al. 2002). In Table 1, we summarize properties of the
displayed shock solutions in the semi-complete space $0<x<+\infty$.

Details of the procedure for finding these semi-complete
self-similar MHD shock solutions will be discussed
presently in section 4.3.
\begin{table*}
\centering
\begin{minipage}{115mm}
\caption{MHD Extensions for Shocks of
Shu et al. (2002) with $\lambda=0.1$.}
\begin{tabular}{|l|c|l|c|c|c|c|c|}
  \hline
  \hline
  description&$A$&$x_{s}$&$v_{d}$&$\alpha_{d}$ &$v_{u}$ &$\alpha_{u}$ \\
  \hline
  $D=11.26$   & 2.00    & 1.42   & 0.532   & 1.59   & 0      & 0.992 \\
  $D=4$       & 1.74    & 1.69   & 0.837   & 1.14   & 0.230  & 0.666 \\
  $D=1.67$    & 1.42    & 1.93   & 1.11    & 0.816  & 0.423  & 0.441 \\
  $D=1$       & 1.20    & 2.06   & 1.27    & 0.653  & 0.534  & 0.337 \\
  $D=2/3$     & 1.02    & 2.15   & 1.38    & 0.537  & 0.620  & 0.269 \\
  $D=10^{-5}$ & $4.30\times10^{-5}$ & 2.56 & 1.91   & $2.10\times10^{-5}$
              & 1.02    & $8.80\times10^{-6}$\\
  $D=10^{-6}$ & $4.30\times10^{-6}$ & 2.56 & 1.91   & $2.10\times10^{-6}$
              & 1.02    & $8.80\times10^{-7}$\\

  \hline
\end{tabular}
\medskip

These solutions can be viewed as Class II similarity MHD shock
solutions matched with asymptotic MHD breezes. Columns 1 to 7
provide relevant parameters for MHD shock solutions: $D$ is
the key parameter in the LP-type asymptotic MHD solutions
(\ref{shockclassIIv}), (\ref{shockclassIIa}) and
(\ref{shockclassIIb}) at small $x$; $A$ is the upstream mass
density parameter; $x_s$ indicates either MHD shock location
or MHD shock speed; $v_{d}$ is the reduced speed downstream of
an MHD shock; $\alpha_d$ is the reduced density downstream of
an MHD shock; $v_u$ is the reduced speed upstream of an MHD
shock; and $\alpha_{u}$ is the reduced density upstream of an
MHD shock.
\end{minipage}
\end{table*}



\subsubsection[]{MHD Extensions for Shock Solutions of Shen \& Lou}

\begin{figure}
 \includegraphics[width=3in]{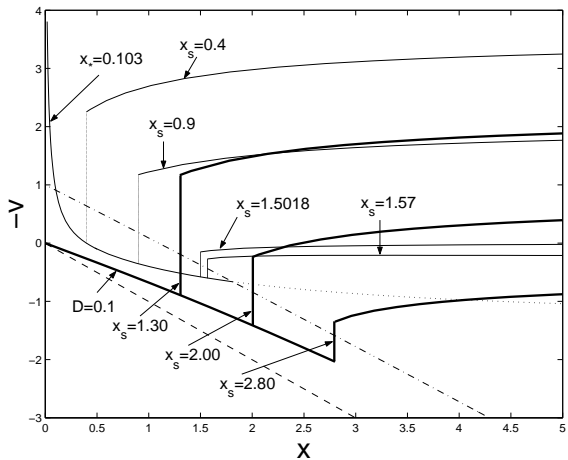}
 \caption{\label{shenlou}
 The first MHD EECC (MEECC) solution with $\lambda=0.1$ crossing the
 magnetosonic critical curve analytically at $x_{*}(1)=0.103$ and
 $x_{*}(2)=1.811$ [within $x_{*}(2)$ is light solid curve, while
 outside $x_{*}(2)$ is dotted curve]. The light solid curves
 represent Class I MHD shock solutions with the downstream as part
 of the first MEECC solution and with shock location $x_{s}$ at
 $0.4$, $0.9$, $1.5018$, $1.57$, respectively; the heavy solid curves
 represent Class II MHD shock solutions with the reduced mass density
 of the central core $D=0.1$ and with the MHD shock location $x_{s}$
 at $1.30$, $2.00$, $2.80$, respectively. Dash-dotted line represents
 magnetosonic critical curve. Dashed line is the mass demarcation line
 $x-v=0$. }
\end{figure}

\begin{figure}
 \includegraphics[width=3in]{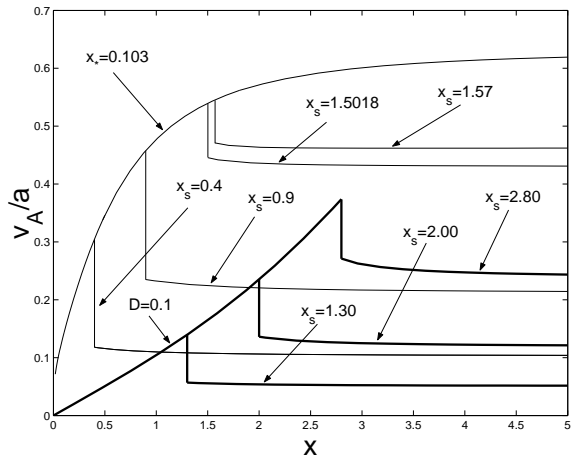}
 \caption{\label{alfvenshenlou}
 The ratio of the Alfv\'en wave speed to the isothermal sound speed
 $v_A/a$ versus $x$ with $\lambda=0.1$ corresponding to the case of
 Fig. \ref{shenlou}. The light solid curves represent Class I MHD
 shock solutions with the downstream as part of the first MEECC
 solution and with the shock location $x_{s}$ at $0.4$, $0.9$,
 $1.5018$, $1.57$, respectively; the heavy solid curves represent
 Class II MHD shock solutions with the reduced mass density of the
 central core $D=0.1$ and with the MHD shock location $x_{s}$ at
 $1.30$, $2.00$, $2.80$, respectively.  }
\end{figure}

\begin{figure}
 \includegraphics[width=3in]{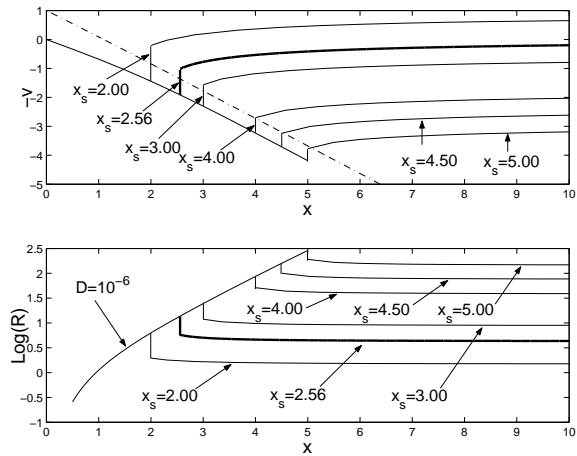}
 \caption{\label{champagne} The case of $\lambda=0.1$.
 The negative reduced radial speed $-v(x)$ (upper panel) and the
 scaled reduced mass denstiy $R(x)\equiv x^{2}\alpha(x)/D$ (lower
 panel) versus $x$. The solid curves form the family of Class II
 MHD shock solutions in the invariant form with the reduced mass
 density of the central core $D\rightarrow 0^{+}$; the heavy solid
 curve is the MHD shock `champagne breeze' solution. Dash-dotted
 curve in the upper panel is the magnetosonic critical curve. }
\end{figure}

\begin{figure}
 \includegraphics[width=3in]{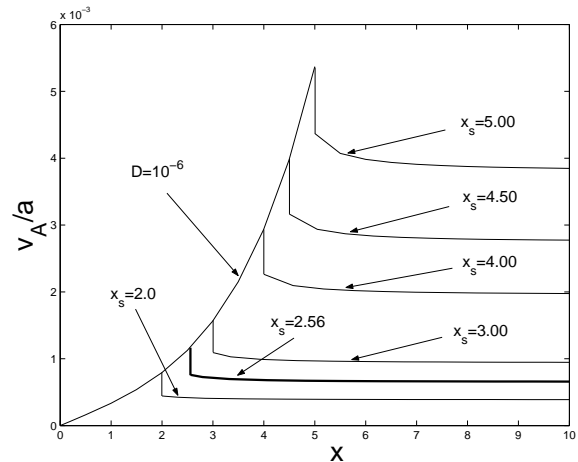}
 \caption{\label{alfvenchampagne}The ratio of the Alfv\'en
 wave speed to the isothermal sound speed $v_A/a$ versus $x$
 with $\lambda=0.1$, corresponding to MHD shock solutions in
 Fig. \ref{champagne}. }
\end{figure}

Shen \& Lou (2004) further extended the Class I shock solutions by
matching with various asymptotic flows and modelled the dynamical
evolution of young stellar objects such as Bok globule B335 system
to account for the observationally inferred mass density and flow
speed profiles as well as the estimated central mass accretion
rate. When a random magnetic field is included, MHD extensions for
this type of shock solution can also be constructed. The
downstream side of such an MHD shock is part of the first type
2-type 1 solution, i.e., the MHD EECC solution (MEECC; Yu \& Lou
2005) which crosses the magnetosonic critical curve analytically.
The two magnetosonic critical points are at $x_{*}(1)=0.103$ and
$x_{*}(2)=1.811$, respectively. By choosing different shock
locations $x_{s}=0.4$, $0.9$, $1.5018$, $1.57$ as in Figure
\ref{shenlou} and Figure \ref{alfvenshenlou}, we readily construct
various upstream MHD solutions as $x\rightarrow+\infty$.
Magnetized `champagne flows' of HII regions around massive OB
stars can also be constructed by allowing for MHD flows at large
$x$. Specific examples of such solutions are presented in Figure
\ref{champagne} and Figure \ref{alfvenchampagne}. Here, the
upstream flows are not necessarily limited to a static magnetized
SIS envelope or magnetized `champagne breezes' with $V=0$. With
different MHD shock locations or speeds, Class II similarity MHD
solutions can be matched with either asymptotic MHD outflows
($V>0$) or asymptotic MHD inflows ($V<0$).

\subsection[]{MHD Similarity Shocks into\\
\ \qquad a Magnetostatic SIS Envelope}

MHD extensions for shock solutions of Tsai \& Hsu (1995) shown
earlier are just two cases of our following similarity MHD shock
solutions into a magnetostatic SIS envelope. In Tsai \& Hsu (1995)
and Shu et al. (2002), two classes of shock solutions were constructed
to match with a static SIS envelope with $V=0$ and $A=2$ or $A<2$. In
Bian \& Lou (2005), hydrodynamic shock solutions of Class I and Class
II were extensively explored to match with a static SIS envelope. One
can also follow the procedure of Hunter to derive the Class I and
Class II MHD shock solutions in a parallel manner. Here, we would
systematically explore the Class I and Class II MHD shock solutions
matched with a magnetostatic SIS envelope by surveying possible
solutions in the speed-density phase diagram of $v$ and $\alpha$. In
Figure \ref{classIglobal}, we present the Class I MHD shock solutions.
In Figure \ref{classIIglobal}, we provide the Class II MHD shock
solutions. In Table 2, we summarize relevant parameters of the
first three MHD shock solutions of both Class I and Class II.

We first consider Class I MHD shock solutions with an outer
magnetostatic SIS envelope, having downstream solutions diverging
towards the centre and upstream solutions of mass parameter $A=2$.
Tsai \& Hsu (1995) obtained a hydrodynamic shock solution for such
a case. Bian \& Lou (2005) systematically extended the shock results
of Tsai \& Hsu (1995) and obtained a wide variety of similarity
shock solutions. We shall fully explore the speed-density phase
diagram for constructing semi-complete MHD shock solutions. For every
assigned value for $x_{*}(1)$ [$x_{*}(1)<x_{m}$ where $x_{m}$ is a
chosen meeting point] and after integrating towards the meeting point
$x_{m}$ with a type 2 MHD eigensolution, we obtain a pair of \{$v$,
$\alpha$\} at the meeting point $x_{m}$ in the so-called phase diagram
of $v$ versus $\alpha$. For a sequence of such $x_{*}(1)$ values, a
series of \{$v$, $\alpha$\} pairs is obtained, giving rise to a curve
in the phase diagram. The upstream is part of a MEWCS. By choosing
an MHD shock location $x_{s}>x_{m}$, we determine the corresponding
upstream values $v_{u}$ and $\alpha_{u}$ at $x_{s}$ along the MEWCS
solution. We then calculate $v_{d}$ and $\alpha_{d}$ using the
isothermal MHD shock conditions in terms of the upstream values
$v_{u}$ and $\alpha_{u}$. Starting from $v_{d}$ and $\alpha_{d}$ at
$x_{s}$, we then integrate the coupled nonlinear MHD ODEs backwards
to the meeting point $x_{m}$ to obtain another pair of \{$v$, $\alpha$\}.
In other words, every value of $x_{s}$ corresponds to a pair of
\{$v$, $\alpha$\}. For a sequence of $x_{s}$ values, another curve
in the phase diagram is thus produced. The intersection points of
the two phase curves represent matches of this type of similarity
MHD shock solutions.

\begin{figure}
 \includegraphics[width=3in]{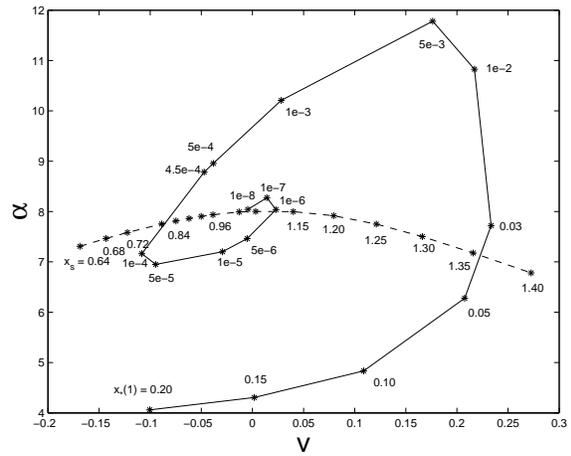}
 \caption{\label{classIphasedia} The phase diagram of Class I
 similarity MHD shock solutions with $\lambda=0.1$. The meeting point
 is chosen at $x_{m}=0.5$. The type 2 MHD eigensolution is used at
 the point $x_{*}(1)$ to integrate towards the meeting point. In
 the range of $x>x_{s}$, the magnetostatic SIS envelope with $A=2$
 and $V=0$ is used to get this diagram. The values of $x_{*}(1)$
 and $x_{s}$ of the first three intersection points are
 $[x_{*}(1)=0.0369,\ x_{s}=1.3599]$, $[x_{*}(1)=2.044\times 10^{-4},
 \ x_{s}=0.7807]$ and $[x_{*}(1)=1.10\times 10^{-4},\ x_{s}=1.1266]$,
 respectively.  }
\end{figure}

\begin{figure}
 \includegraphics[width=3in]{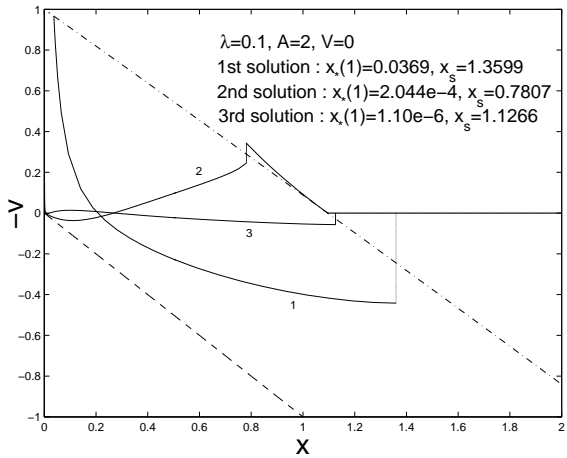}
 \caption{\label{classIglobal}
 The first three Class I MHD shock solutions with $\lambda=0.1$.
 The MHD shock solution 2 is an accretion MHD shock. The reduced radial
 speeds of the preshock and postshock are both negative. This similarity
 accretion MHD shock expands at a constant sub-magnetosonic speed. The
 type 2 MHD eigensolution is chosen at the point $x_{*}(1)$ to integrate
 towards the meeting point. In the range of $x>x_{s}$, the magnetostatic
 SIS envelope with $A=2$ and $V=0$ is used to get this diagram. The
 dash-dotted curve represents the magnetosonic critical curve. The
 dashed line is $x-v=0$. The corresponding speed ratio $v_A/a$ is
 shown in Fig. \ref{alfvenclassIglobal}. }
\end{figure}

\begin{figure}
 \includegraphics[width=3in]{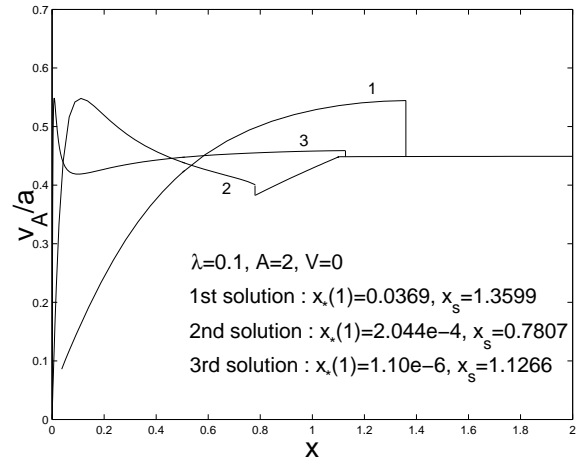}
 \caption{\label{alfvenclassIglobal}
 Corresponding to the shock solutions in Fig. \ref{classIglobal},
 the ratio of the Alfv\'en wave speed to the isothermal sound speed
 $v_A/a$ versus $x$ with $\lambda=0.1$. The first three Class I MHD
 shock solutions identified in the $\alpha-v$ phase diagram of Fig.
 \ref{classIphasedia}. The type 2 MHD eigensolution is taken at the
 point $x_{*}(1)$ to integrate towards the meeting point. In the
 range of $x>x_{s}$, the magnetostatic SIS envelope with $A=2$
 and $V=0$ is used for these MHD shocks. }
\end{figure}

\begin{figure}
\begin{center}
\includegraphics[scale=0.43]{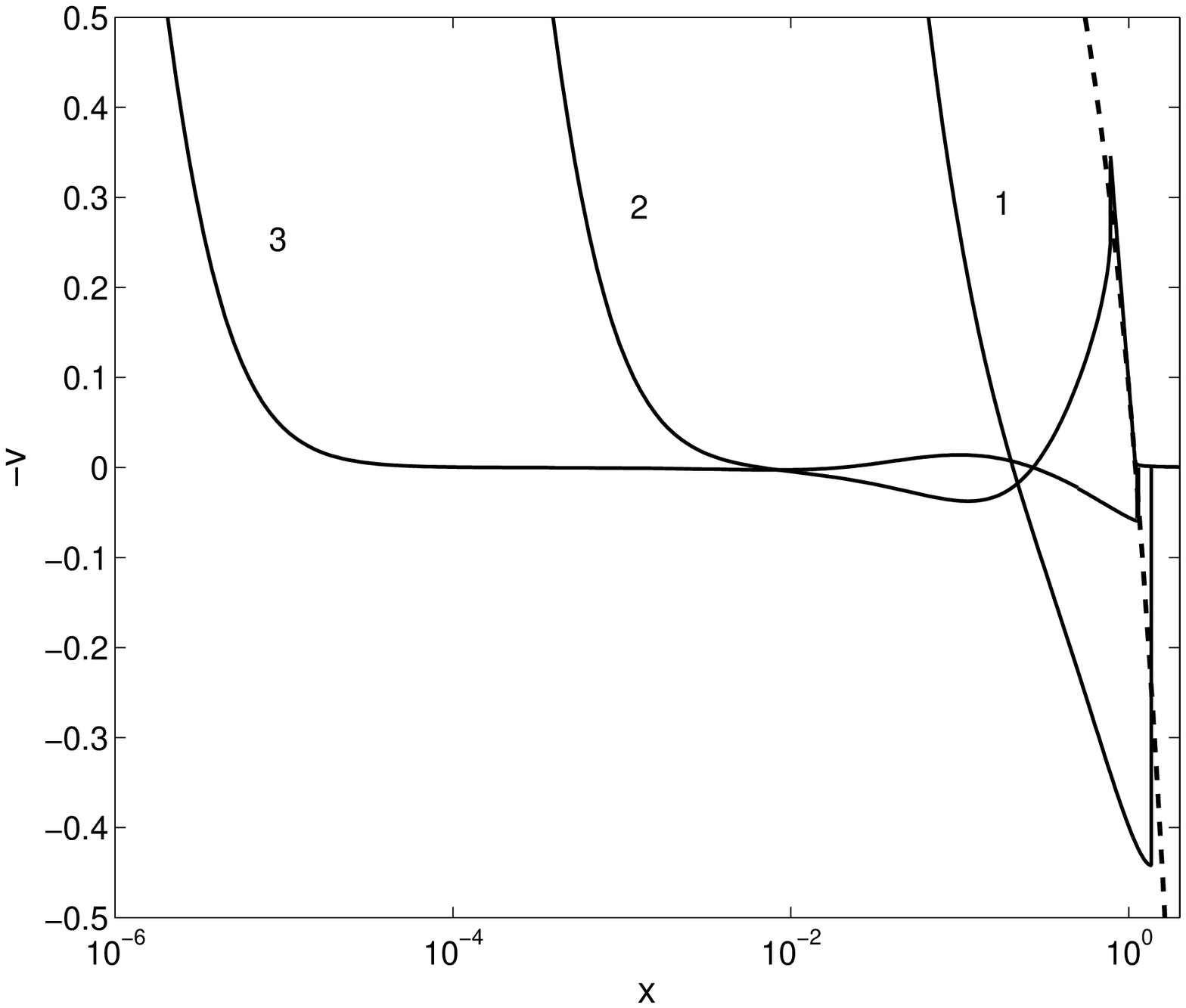}
\caption{Enlarged portions of Class I MHD shock solution curves
near $x\rightarrow 0^{+}$ of Fig.~\ref{classIglobal}, emphasizing
the diverging and oscillatory behaviours of this family of MHD
shock solutions as $x\rightarrow0^{+}$. The $x-$axis is shown
in the logarithmic scale. The dashed line on the right is the
magnetosonic critical curve. The undulatory profiles of the curves
marked by numerals 1, 2, 3 represent self-similar sub-magnetosonic
radial oscillations with one, two, three stagnation points,
respectively. }
\label{classIenlarge}
\end{center}
\end{figure}

In Fig. \ref{classIphasedia}, we show the relevant phase diagram of
$v$ and $\alpha$ following the matching procedure described above
for a chosen meeting point at $x_{m}=0.5$. Relevant parameters of
the first three intersection points in the phase diagram are
($x_{*}(1),x_{s}$)=(0.0369, 1.3599), ($2.044\times 10^{-4}, 0.7807$),
($1.10\times 10^{-4}, 1.1266$), respectively.
In Figs. \ref{classIglobal} and \ref{alfvenclassIglobal},
we display the first three global MHD shock solutions of this
type. Enlarged portions of these solutions are given in Fig
\ref{classIenlarge}.

\begin{figure}
 \includegraphics[width=3in]{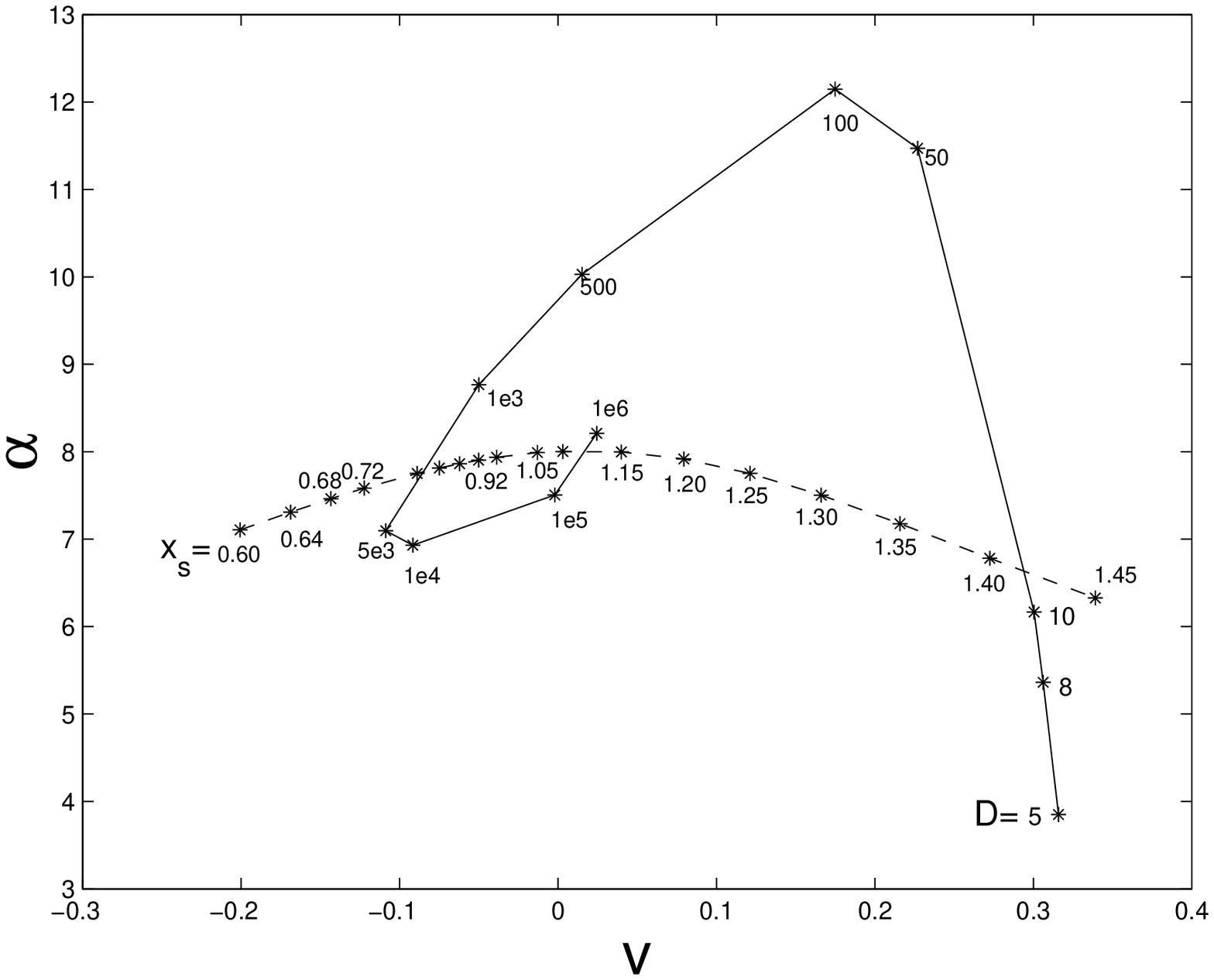}
 \caption{\label{classIIphasedia} The phase diagram of Class II
 similarity MHD shock solutions with $\lambda=0.1$. The meeting
 point is chosen at $x_{m}$=0.5. The MHD LP type solution is used
 at the point near origin to integrate outward to the meeting point.
 In the range of $x>x_{s}$, the magnetostatic SIS envelope with $A=2$
 and $V=0$ is specified for this diagram. The values of $x_{s}$ and
 $D$ of the first three intersection points are $(x_{s}=1.41944,\
 D=11.26)$, $(x_{s}=0.77598,\ D=2156)$ and $(x_{s}=1.12668,\
 D=4.04\times 10^{5})$, respectively.  }
\end{figure}

\begin{figure}
 \includegraphics[width=3in]{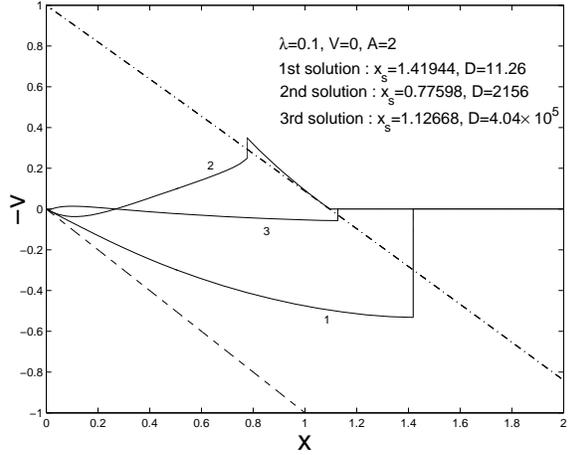}
 \caption{\label{classIIglobal}
 The first three class II MHD shock solutions with $\lambda=0.1$
 identified in the phase diagram of Fig. \ref{classIIphasedia}.
 The shock solution 2 is an accretion MHD shock with the reduced radial
 speeds of the preshock and postshock being both negative. This accretion
 MHD shock front expands at a constant sub-magnetosonic speed. The LP
 type MHD solution is used at near the origin to integrate outward to the
 meeting point. In the range of $x>x_{s}$, the magnetostatic SIS envelope
 with $A=2$ and $V=0$ is specified for this diagram. The dash-dotted line
 represents the magnetosonic critical curve and the straight dashed line
 represents the demarcation line of $x-v=0$ for a positive enclosed mass.
 The speed ratio of the Alfv\'en wave speed to the isothermal sound
 speed $v_A/a$ versus $x$ is shown in Fig. \ref{alfvenclassIIglobal}. }
\end{figure}

\begin{figure}
\includegraphics[width=3in]{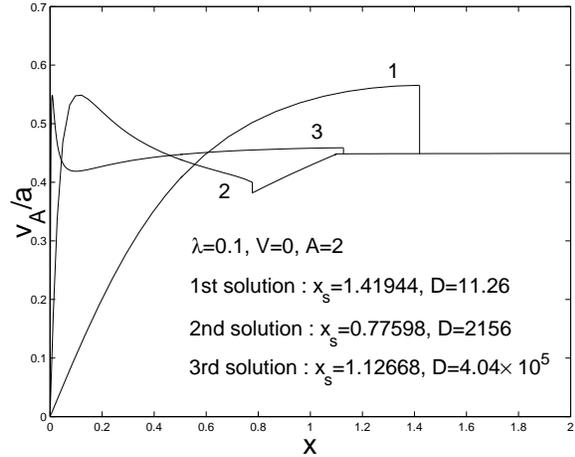}
\caption{\label{alfvenclassIIglobal} For the first three Class II MHD
shock solutions with $\lambda=0.1$ identified in the phase diagram of
Fig. \ref{classIIphasedia} and shown in Fig. \ref{classIIglobal}, the
corresponding ratio of the Alfv\'en wave speed to the isothermal sound
speed $v_A/a$ versus $x$ is in display here. The MHD LP type solution
is used near the origin to integrate outward to the meeting point. In
the range of $x>x_{s}$, the magnetostatic SIS envelope with $A=2$ and
$V=0$ is specified for this diagram.
}
\end{figure}

When it comes to the Class II MHD shock solutions that connect
downstream MHD LP solutions with a magnetostatic SIS envelope, the
method of finding the solutions is similar to those described
above. The key difference is that the downstream is now replaced
by the MHD LP type solutions. In parallel, we could obtain similar
phase diagram to identify the intersection points of phase curves.
In Fig. \ref{classIIphasedia}, we present a relevant phase diagram
of $v$ and $\alpha$ to match similarity MHD shock solutions. In
Fig. \ref{classIIglobal} and Fig. \ref{alfvenclassIIglobal}, we
display the first three global MHD shock solutions of this type.

In Table 2, we summarize the properties of Class I and Class II
MHD shock solutions matched with a magnetostatic SIS envelope,
i.e., MEWCS solutions.

\begin{table*}
\centering
\begin{minipage}{115mm}
\caption{Semi-complete Class I and II similarity MHD shock
solutions matched with a magnetostatic SIS envelope. }
\begin{tabular}{|l|l|l|c|l|l|l|l|}
  \hline
  \hline
  Description&$m_{0}$&$x_{s}$&$v_{d}$&$\alpha_{d}$&$v_{u}$
   &$\alpha_{u}$&N\\
  \hline
  $x_{*}(1)=0.0369$ &0.0726  &1.3599&0.4421 &1.6024&0&1.0815&1\\
  $x_{*}(1)=2.044\times 10^{-4}$&$4.09\times 10^{-4}$&0.7807&
        -0.2450&2.6333&-0.3428&2.4019&2\\
  $x_{*}(1)=1.10\times 10^{-6}$&$2.20\times 10^{-6}$
           &1.1266&0.0567&1.6593&0&1.5758&3\\
  \hline
  $D=11.26$&-&1.41944&0.5318 &1.5847&0&0.9926&0\\
  $D=2156$&-&0.77598&-0.2478&2.6593&-0.3501&2.4177&1\\
  $D=4.04\times 10^{5}$&-&1.12668&0.0569 &1.6593&0&1.5755&2\\
  \hline
\end{tabular}
\medskip

Columns 1 to 8 provide relevant parameters for MHD shock solutions:
$x_{*}(1)$ is the inner magnetosonic critical point for Class I MHD
solutions and $D$ is the central `density parameter' of MHD LP type
symptotic solution for constructing Class II MHD solutions; $m_{0}$
is the reduced central mass accretion rate; $x_{s}$ corresponds to
both the shock location or shock speed; $v_{d}$ is the downstream
reduced speed; $\alpha_{d}$ is the downstream reduced density; $v_{u}$
is the upstream reduced speed; $\alpha_{d}$ is the upstream reduced
density and the number of stagnation points $N$ indicates the number
of nodes for sub-magnetosonic self-similar oscillations in MHD shock
solutions.
\end{minipage}
\end{table*}

\subsection[]{Similarity MHD Shock Breezes}

\begin{figure}
 \includegraphics[width=3in]{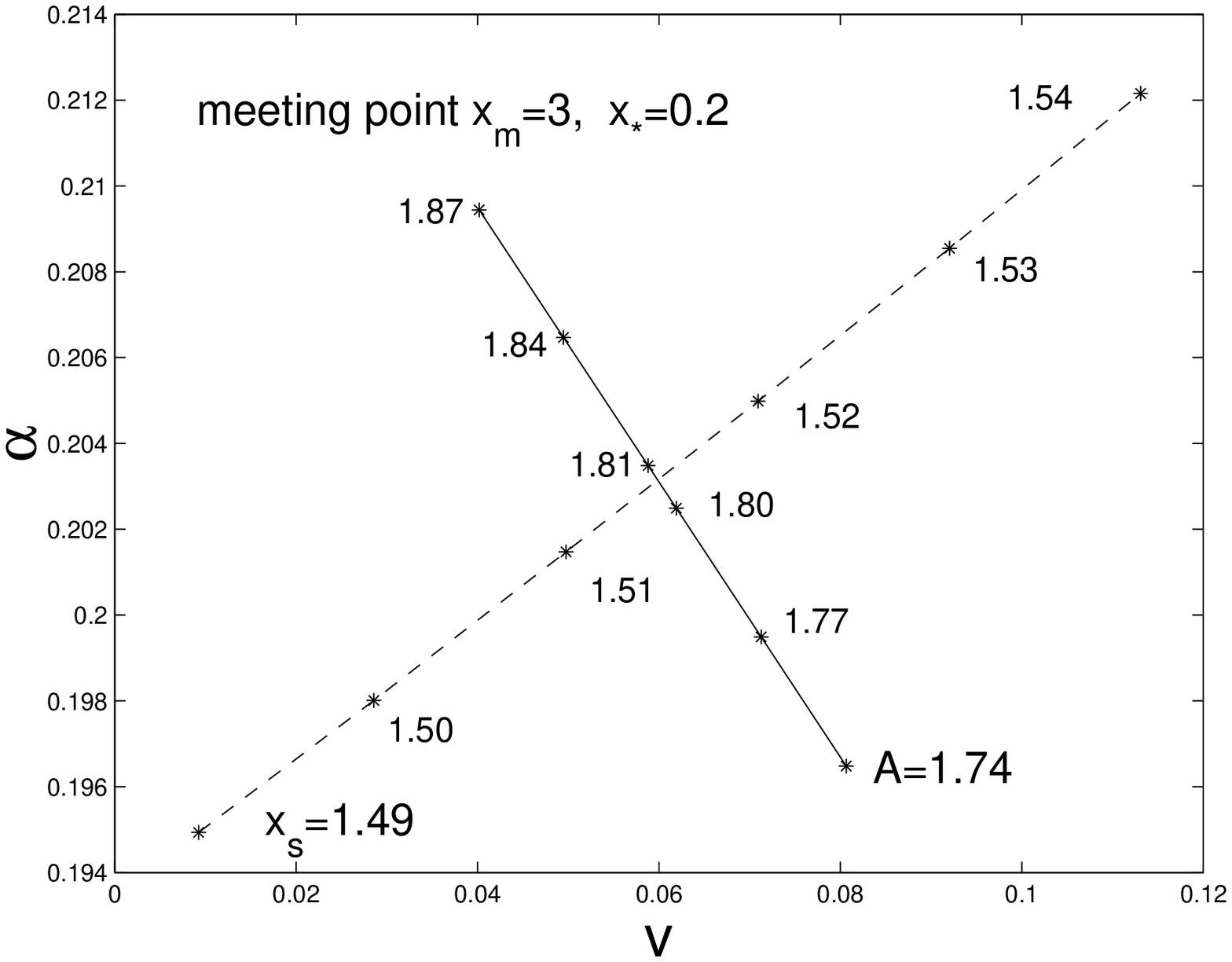}
 \caption{\label{p2phasedia} An example of illustration for the phase
 diagram of Class I MHD shock breeze solution with $\lambda=0.1$ for
 the case of $x_{*}=0.2$. For other values of $x_{*}$, similar phase
 diagrams are necessary to identify the relevant global MHD shock
 solutions. The type 2 MHD eigensolution is specified at the point
 $x_{*}(1)$ to integrate towards the meeting point. Analytical
 asymptotic MHD solution (i) as given by equations
 $(\ref{shockinftyv})-(\ref{shockinftyb})$ with $V=0$ and a varying $A$
 parameter is specified to integrate also towards the meeting point.
}
\end{figure}

\begin{figure}
 \includegraphics[width=3in]{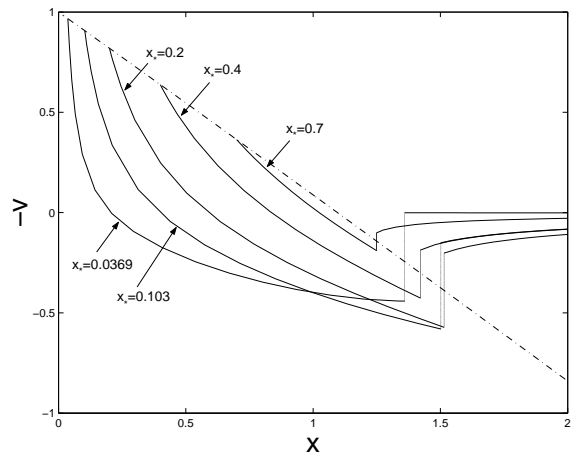}
 \caption{\label{classIbreeze}
 The Class I MHD shock solutions with $\lambda=0.1$.
 The type 2 MHD eigensolution is specified at the point $x_{*}(1)$
 along the magnetosonic critical curve to integrate towards the
 meeting point $x_m$. Analytical asymptotic MHD solution (i) as
 given by equations $(\ref{shockinftyv})-(\ref{shockinftyb})$
 with $V=0$ and a varying $A$ parameter is invoked to also
 integrate towards the meeting point $x_m$. The dash-dotted
 line represents the magnetosonic critical curve. The speed
 ratio of $v_A/a$ is shown in Fig. \ref{alfvenclassIbreeze}.  }
\end{figure}

\begin{figure}
 \includegraphics[width=3.5in]{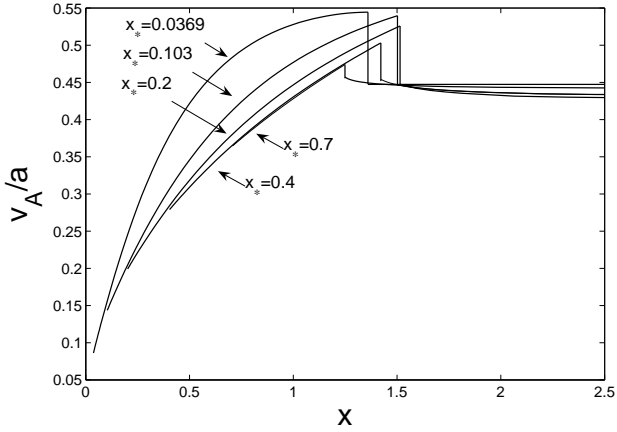}
 \caption{\label{alfvenclassIbreeze}
 The ratios of the Alfv\'en wave speed and the isothermal
 sound speed $v_A/a$ with $\lambda=0.1$ corresponding to
 the MHD shock breeze results in Fig. \ref {classIbreeze}. }
\end{figure}

MHD extensions for shock solutions of Shu et al. (2002) described
in previous section should be viewed as a subset of Class II type
MHD shock solutions. We now consider Class I type MHD shock
solutions and allow the mass parameter $A$ to vary (i.e., $A$
cannot be equal to 2 in general) for a given $x_{*}$ corresponding
to a specific reduced central mass accretion rate $m_{0}$. Both
parameters $A$ and $x_{s}$ are adjusted gradually to match the
upstream and downstream solutions in the phase diagram of $v$
versus $\alpha$. We integrate from a specified $x_{*}$ along the
magnetosonic curve (in our calculations, we choose $x_{*}=0.0369,\
0.103,\ 0.2,\ 0.4,\ 0.7$ as several examples of illustration) with
a type 2 MHD eigensolution towards the shock front at $x_{s}$ to
obtain the two parameters $v_{d}$ and $\alpha_{d}$. By the
isothermal MHD shock conditions, we determine $v_{u}$ and
$\alpha_{u}$ for a given pair of $v_{d}$ and $\alpha_{d}$.
Starting from $v_{u}$ and $\alpha_{u}$ at $x_{s}$, we then
integrate further the coupled nonlinear MHD ODEs to a chosen
meeting point $x_{m}=3.0$. Here, we apply the analytical
asymptotic MHD solution (i) as given by equations
$(\ref{shockinftyv}) -(\ref{shockinftyb})$ at $x=20$ as the
far-away `boundary condition' to integrate backwards to
$x_{m}=3.0$. For a given $x_{*}$ value and two series of $x_{s}$
and $A$ values, we then draw a relevant phase diagram to determine
the intersection point of the two phase curves, i.e., $x_{s}$ and
$A$. We could then construct such a similarity MHD shock breeze
using the values of $x_{s}$ and $A$ at the intersection point. In
Fig. \ref{p2phasedia}, we display a sample phase diagram for the
case of $x_{*}=0.2$. For $x_{*}=0.0369,\ 0.103,\ 0.4,\ 0.7$
respectively, we can obtain qualitatively similar phase diagrams
(not shown here to avoid redundancy) and their corresponding
$x_{s}$ and $A$ values. In Fig. \ref{classIbreeze}, we present
Class I isothermal MHD shock breeze solutions. The corresponding
ratios of the Alfv\'en wave speed $v_A$ to the isothermal sound
speed $a$ are shown in Fig \ref{alfvenclassIbreeze}. Table 3
contains the major relevant parameters for the shock solution
examples of illustration.

Methods of determining Class II solutions of MHD shock breeze are
similar to the above procedure. The main difference is to adopt the
MHD LP type solutions for the downstream. Results of Class II MHD
shock solutions have been discussed already in subsection 4.1.2
and we shall not repeat here.

\begin{table*}
\centering
\begin{minipage}{115mm}
\caption{Semi-complete Class I similarity MHD shock
solutions matched with an asymptotic MHD breeze.}
\begin{tabular}{|l|c|l|c|c|l|l|l|l|}
  \hline
  \hline
  $x_{*}$&$m_{0}$&$A$&$x_{s}$&$v_{d}$&$\alpha_{d}$&$v_{u}$&$\alpha_{u}$\\
  \hline
  $x_{*}=0.0369$& 0.0724&2     &1.3599&0.4414&1.6020&0     &1.0828\\
  $x_{*}=0.103$ & 0.1952&1.8530&1.5018&0.5805&1.2907&0.1507&0.8802\\
  $x_{*}=0.2$   & 0.3586&1.8064&1.5146&0.5718&1.2050&0.2020&0.8655\\
  $x_{*}=0.4$   & 0.6356&1.8514&1.4216&0.4265&1.2520&0.1870&1.0091\\
  $x_{*}=0.7$   & 0.9107&1.9484&1.2492&0.1897&1.4401&0.1014&1.3294\\
  \hline
\end{tabular}
\medskip

Columns 1 to 8 provide the relevant parameters for MHD shock solutions:
$x_{*}$ is the magnetosonic critical point for Class I MHD shock breeze
solutions; $m_{0}$ corresponds to the reduced central mass accretion rate;
$A$ is the mass density parameter of the upstream; the shock location is
at $x_{s}$; $v_{d}$ is the downstream reduced velocity; $\alpha_{d}$ is
the downstream reduced density; $v_{u}$ is the upstream reduced velocity;
and $\alpha_{u}$ is the upstream reduced density.
\end{minipage}
\end{table*}

\subsection[]{Self-Similar Twin MHD Shock Solutions}

\begin{figure}
 \includegraphics[width=3in]{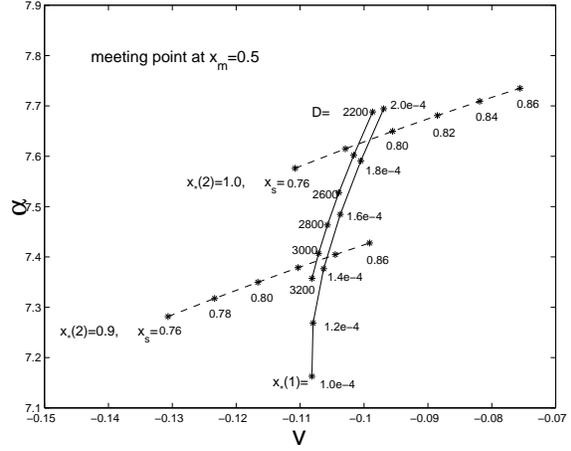}
 \caption{\label{twinphasedia}
 A phase diagram of $v$ versus $\alpha$ at a chosen meeting
 point $x_{m}=0.5$ for constructing MHD shock solutions
 with $\lambda=0.1$. In this phase diagram, both cases of
 $x_{*}(2)=0.9$ and $x_{*}(2)=1.0$ for Class I MHD shock
 solutions and of $x_{*}=0.9$ and $x_{*}=1.0$ for Class II
 MHD shock solutions are shown.  }
\end{figure}

\begin{figure}
 \includegraphics[width=3in]{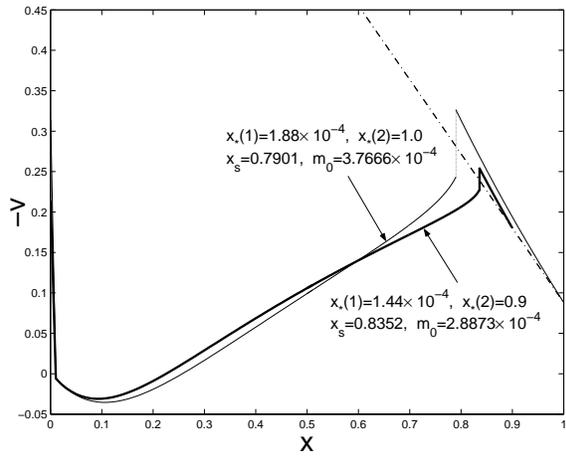}
 \caption{\label{classItwin} Class I MHD shock solutions with
 $\lambda=0.1$ for $x_{*}(2)=0.9$ and $1.0$ shown in terms of
 $-v(x)$ versus $x$ by the heavy and light solid curves,
 respectively. The two corresponding values of inner
 magnetosonic point $x_{*}(1)$ are $1.44\times 10^{-4}$ and
 $1.88\times 10^{-4}$, respectively. The two corresponding
 inner MHD shocks are located at $x_{s}=0.8352$ and
 $x_{s}=0.7901$, respectively. The dash-dotted line
 represents the magnetosonic critical curve. The
 speed ratio $v_A/a$ versus $x$ is shown in Fig.
 \ref {alfvenclassItwin}.  }
\end{figure}

\begin{figure}
 \includegraphics[width=3in]{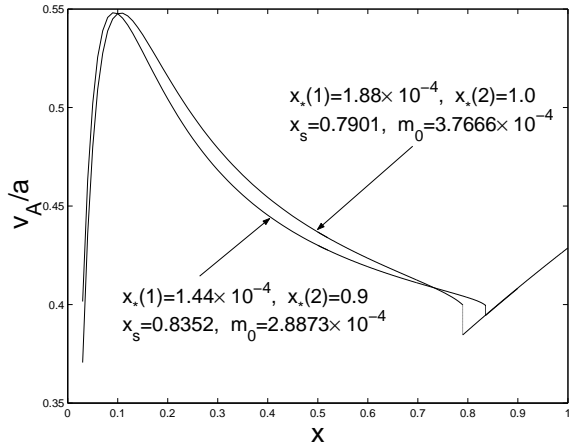}
 \caption{\label{alfvenclassItwin}
 Class I MHD shock solutions with $\lambda=0.1$ for $x_{*}(2)=0.9$
 and $1.0$ for the ratio of the Alfv\'en wave speed and the
 isothermal sound speed $v_A/a$ versus $x$ for the solutions shown
 in Fig. \ref{classItwin}. The two corresponding values of $x_{*}(1)$
 are $1.44\times 10^{-4}$ and $1.88\times 10^{-4}$, respectively. The
 corresponding inner MHD shock is located at $x_{s}=0.8352$ and
 $x_{s}=0.7901$, respectively.  }
\end{figure}

In parallel with the investigation of Bian \& Lou (2005), we can
also construct twin shock solutions in the presence of a random
magnetic field. All MHD shock solutions shown so far contain just
a single MHD shock. In this subsection, we are going to explore
examples of semi-complete similarity solutions with twin MHD
shocks. The upstream between the outer MHD shock and the
magnetostatic state are tangent to the magnetosonic critical curve
at the point of $x_{*}(2)=(1+2\lambda)^{1/2}$, which actually is
part of MEWCS (Yu \& Lou 2005). We may shift the location
$x_{*}(2)$ somewhat towards the origin into the region of
$0<x<(1+2\lambda)^{1/2}$ and the upstream solution of the inner
MHD shock may cross the magnetosonic critical curve analytically
with a type 2 MHD eigensolution. We can subsequently construct
Class I and Class II twin MHD shock solutions in the $\alpha-v$
phase diagram. For constructing Class I twin MHD shock solutions,
we first specify the value of $x_{*}(2)$ and adjust the value of
$x_{*}(1)$ point and the shock location $x_{s}(1)$ between
$x_{*}(1)$ and $x_{*}(2)$ to search for similarity MHD shock
solution crossing the magnetosonic critical curve analytically
using the intersections of phase curves in the $\alpha-v$ phase
diagram. More specifically, we choose a meeting point $x_{m}=0.5$
and integrate an upstream solution of the inner shock from
$x_{*}(2)$ for Class I solutions using a type 2 MHD eigensolution
as initial conditions to a chosen shock location $x_{s}(1)$ in
order to get the upstream values of $v_{u}$ and $\alpha_{u}$ at
the inner shock location $x_{s}(1)$. We next use the isothermal
MHD shock conditions to determine the downstream values of $v_{d}$
and $\alpha_{d}$ at the inner shock location $x_{s}(1)$. We then
integrate from $v_{d}$ and $\alpha_{d}$ at $x_{s}(1)$ towards the
meeting point $x_{m}=0.5$. Meanwhile, we also integrate forward
from $x_{*}(1)$ again using type 2 MHD eigensolution to
$x_{m}=0.5$. In Fig. \ref{twinphasedia}, we show the phase diagram
of Class I MHD shock solution for two cases $x_{*}(2)=0.9$ and
$1.0$, respectively (n.b., phase curves of Class II MHD shock
solution are also included in this figure). From the intersections
of phase curves, we obtain the relevant parameter pair [$x_{*}(1)$,
$x_{s}(1)$] of Class I MHD shock solutions as ($1.44\times 10^{-4},
\ 0.8352$) for the case of $x_{*}(2)=0.9$ and as ($1.88\times 10^{-4},
\ 0.7901$) for the case of $x_{*}(2)=1.0$. We show the corresponding
Class I MHD shock solutions for $-v$ versus $x$ in Fig. \ref{classItwin}
as an example of illustration. In Fig. \ref{alfvenclassItwin}, we
show the corresponding ratios of the Alfv\'en wave speed $v_A$ to
the isothermal sound speed $a$.

\begin{figure}
 \includegraphics[width=3in]{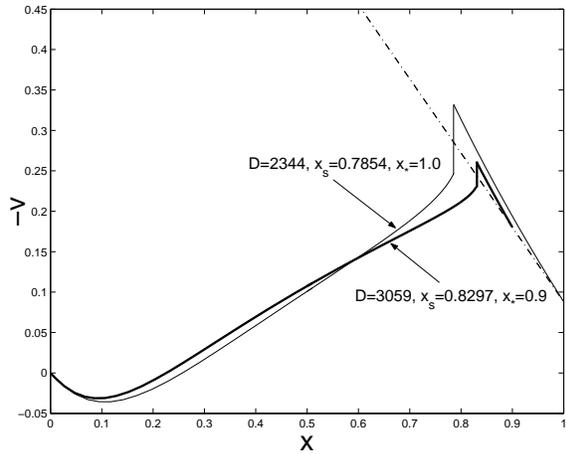}
 \caption{\label{classIItwin}
 Class II MHD shock solutions with $\lambda=0.1$ for $x_{*}=0.9$
 and $1.0$ in terms of $-v(x)$ versus $x$ by the heavy and light
 solid curves, respectively. The corresponding values of
 $(D,\ x_{s})$ are $(3059,\ 0.8297)$ and $(2344,\ 0.7854)$,
 respectively. The dash-dotted line represents the
 magnetosonic critical curve. The speed ratio $v_A/a$
 versus $x$ is shown in Fig. \ref{alfvenclassIItwin}. }
\end{figure}

\begin{figure}
 \includegraphics[width=3in]{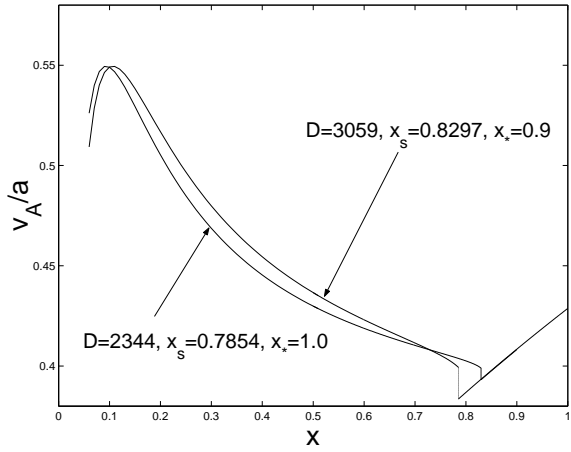}
 \caption{\label{alfvenclassIItwin} Class II MHD shock solutions
 with $\lambda=0.1$ for $x_{*}=0.9$ and $1.0$ shown for the ratio
 of the Alfv\'en wave speed to the isothermal sound speed $v_A/a$
 versus $x$ corresponding to the results of Fig. \ref{classIItwin}.
 The corresponding values of ($D,\ x_{s}$) are $(3059,\ 0.8297)$
 and $(2344,\ 0.7854)$, respectively. }
\end{figure}

For constructing Class II twin MHD shock solutions, we first specify
the value of $x_{*}$ and adjust the value of central reduced mass
density $D$ parameter together with the MHD shock location $x_{s}(1)$
between $0$ and $x_{*}$ to search for similarity MHD shock solutions
using the intersections of phase curves in the $\alpha-v$ phase diagram.
From the intersections of phase curves in Fig. \ref{twinphasedia}, we
identify the relevant parameter pair [$D,\ x_{s}(1)$] of Class II MHD
shock solutions as $(3059,\ 0.8297)$ for the case of $x_{*}=0.9$ and
as $(2344,\ 0.7854)$ for the case of $x_{*}=1.0$. We present the
corresponding Class II MHD shock solutions for $-v$ versus $x$ in Fig.
\ref{classIItwin} as another example of illustration. The corresponding
ratios of the Alfv\'en wave speed $v_A$ to the isothermal sound speed
$a$ are shown in Fig \ref{alfvenclassIItwin}.

In Fig. \ref{twinphasedia}, relevant phase curves of both Class
I and Class II twin MHD shock solutions are presented, where the
meeting point $x_{m}=0.5$ is chosen. In this phase diagram, both
cases of $x_{*}(2)=0.9$ and $x_{*}(2)=1.0$ for Class I solution
and of $x_{*}=0.9$ and $x_{*}=1.0$ for Class II solution are
plotted. Using the intersection points of phase curves in this
phase diagram, we can readily construct the inner Class I and
Class II MHD shock solutions, respectively.

\begin{figure}
 \includegraphics[width=3in]{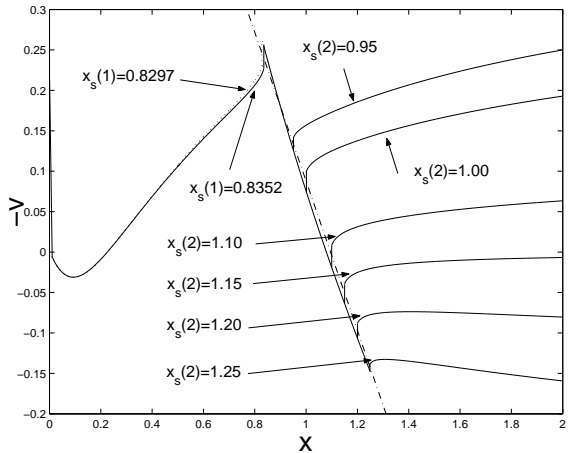}
 \caption{\label{twinshockp9}
 Class I (solid curve) and Class II (dotted curve) twin MHD shock
 solutions with $\lambda=0.1$ for $x_{*}=0.9$ are shown in terms
 of $-v(x)$ versus $x$. The dash-dotted line represents the
 magnetosonic critical curve. The outer MHD shock locations
 $x_{s}(2)$ are located at $0.95,\ 1.00,\ 1.10,\ 1.15,\ 1.20,\
 1.25$ to match with various asymptotic MHD solutions at
 $x\rightarrow+\infty$. The corresponding speed ratios $v_A/a$
 versus $x$ are shown in Fig. \ref{alfventwinshockp9}. }
\end{figure}

\begin{figure}
 \includegraphics[width=3in]{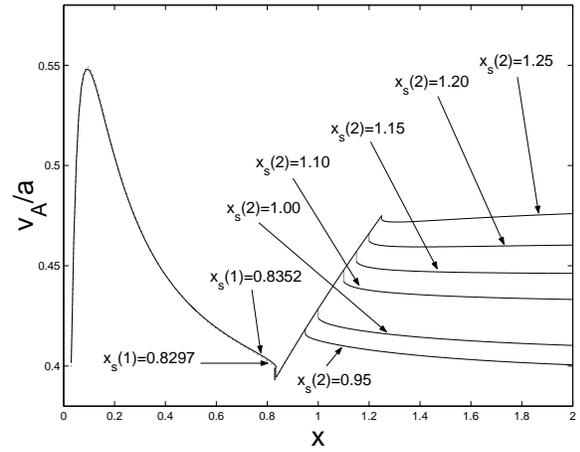}
 \caption{\label{alfventwinshockp9} Class I (solid curve) and Class
 II (dotted curve) twin MHD shock solutions with $\lambda=0.1$
 for $x_{*}=0.9$ are shown in terms of the ratio of the Alfv\'en
 wave speed to the isothermal sound speed $v_A/a$ versus $x$
 corresponding to results of Fig. \ref{twinshockp9}.
 The outer MHD shocks locations $x_{s}(2)$ are at $0.95,\ 1.00,\
 1.10,\ 1.15,\ 1.20,\ 1.25$ respectively to match with various
 asymptotic MHD solutions at $x\rightarrow+\infty$.  }
\end{figure}

\begin{figure}
 \includegraphics[width=3in]{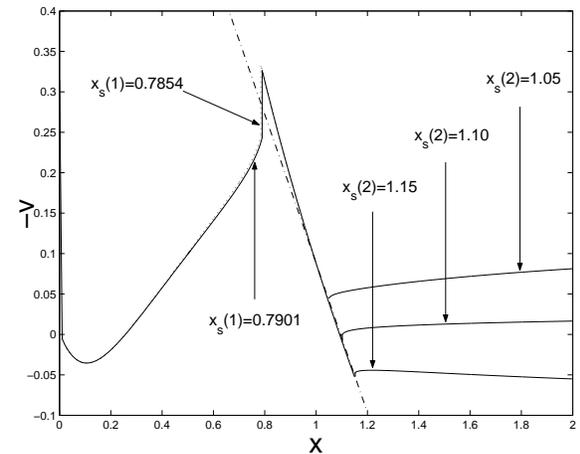}
 \caption{\label{twinshockone} Class I (solid curve) and Class
 II (dotted curve) twin MHD shock solutions with $\lambda=0.1$
 and the magnetosonic point $x_{*}=1.0$ shown in terms of
 $-v(x)$ versus $x$. The dash-dotted line represents the
 magnetosonic critical curve. The outer MHD shock locations
 $x_{s}(2)$ are at $1.05,\ 1.10,\ 1.15$, respectively to match
 with various asymptotic MHD solutions at $x\rightarrow+\infty$.
 The corresponding speed ratio $v_A/a$ versus $x$ is shown in
 Fig. \ref{alfventwinshockone}. }
\end{figure}

\begin{figure}
 \includegraphics[width=3in]{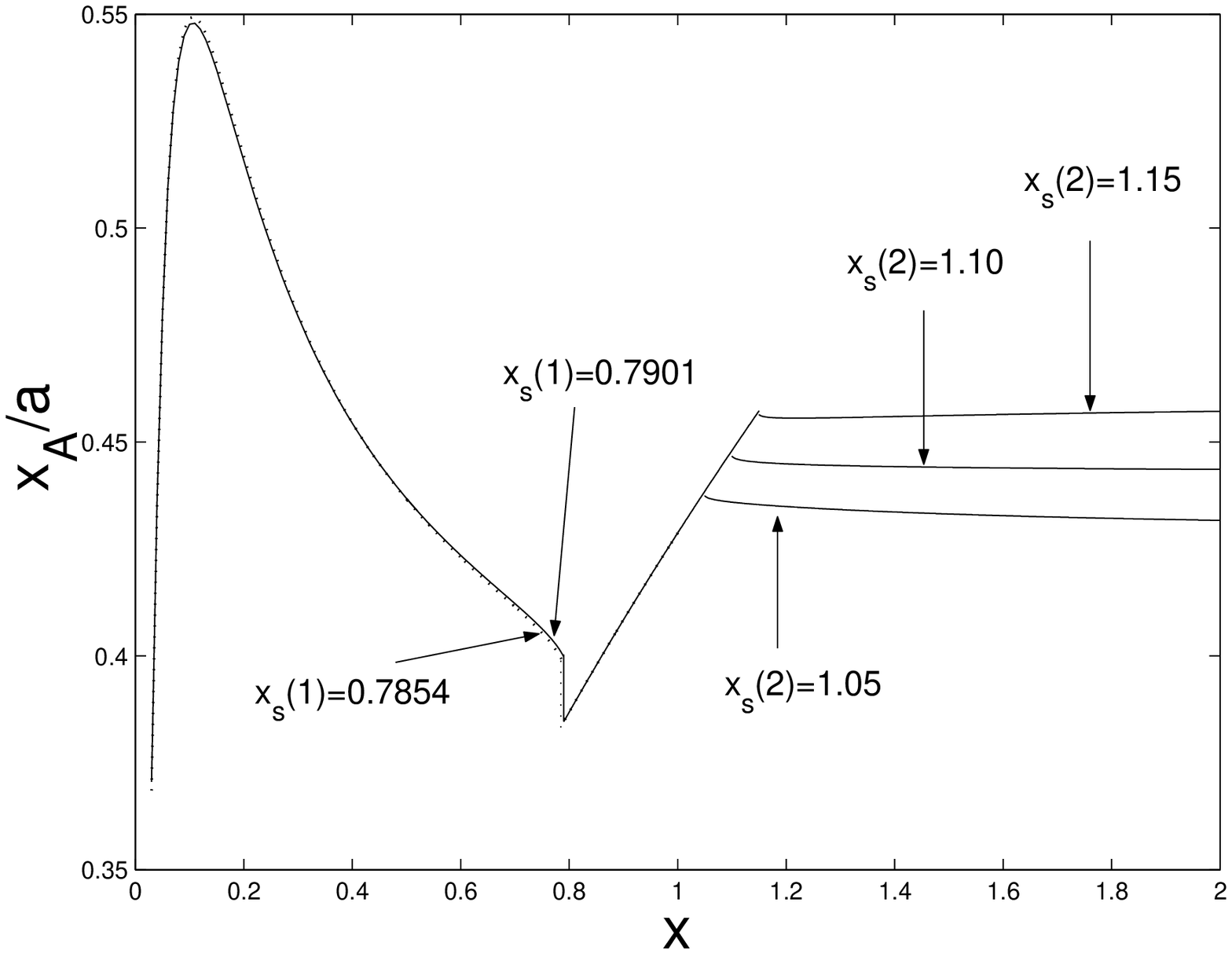}
 \caption{\label{alfventwinshockone}
 Class I (solid curve) and Class II (dotted curve) twin MHD shock
 solutions with $\lambda=0.1$ and a magnetosonic critical point
 $x_{*}=1.0$ shown in terms of the speed ratio $v_A/a$ versus $x$
 corresponding to the MHD shock results of Fig. \ref{twinshockone}.
 The outer MHD shock locations $x_{s}(2)$ are at $1.05,\
 1.10,\ 1.15$, respectively, to match with various asymptotic
 MHD solutions at $x\rightarrow+\infty$.  }
\end{figure}

The above procedures are implemented to search for the inner MHD
shock solutions. In the range of $x>x_{*}(2)$ for Class I solution
(or $x>x_{*}$ for Class II solution), we can further construct the
outer MHD shock by choosing different outer shock location
$x_{s}(2)$. For $x_{*}(2)=0.9$ (for Class I) or $x_{*}=0.9$ (for
Class II), we choose $x_{s}(2)=0.95,\ 1.00,\ 1.10,\ 1.15,\ 1.20, \
1.25$, respectively, while for $x_{*}(2)=1.0$ (for Class I) or
$x_{*}=1.0$ (for Class II), we choose $x_{s}(2)=1.05,\ 1.10,\
1.15$, respectively. The upstream asymptotic MHD solutions across
the outer shock can match with the analytical asymptotic solution
$(\ref{shockinftyv})-(\ref{shockinftyb})$ for
$x\rightarrow+\infty$. When the value of the outer shock location
$x_{s}(2)$ is smaller, the upstream would be an MHD inflow
(contraction or accretion). As the value of MHD shock location
$x_{s}(2)$ increases, the upstream would become an MHD outflow
(expansion or wind or breeze). Relevant solutions are displayed in
Fig. \ref{twinshockp9}$-$Fig. \ref{alfventwinshockone}

\begin{table*}
\centering
\begin{minipage}{115mm}
\caption{Semi-complete Class I and II similarity
twin MHD shock solutions matched with different
outer magnetosonic critical point $x_{*}(2)$. }
\begin{tabular}{|l|l|l|c|r|r|}
  \hline
  \hline
  $x_{*}(2)\quad$ & Description & $\quad x_{s}(1)\qquad$
  & $\quad x_{s}(2)\qquad$ & $\quad A\qquad$ &$V\qquad$\\
  \hline
  0.90&$x_{*}(1)=1.44\times 10^{-4}$&0.8352&0.95&1.5552&-0.3852\\
      &$m_{0}=2.8873\times 10^{-4}$ &      &    &    &         \\
      &$D=3059$                     &0.8297&1.00&1.6398&-0.3034\\
      &                             &      &1.10&1.8546&-0.1103\\
      &                             &      &1.15&1.9868& 0     \\
      &                             &      &1.20&2.1401& 0.1198\\
      &                             &      &1.25&2.3238& 0.2553\\
  \hline
  1.00&$x_{*}(1)=1.88\times 10^{-4}$&0.7901&1.05&1.8414&-0.1298\\
      &$m_{0}=3.7666\times 10^{-4}$ &      &    &      &       \\
      &$D=2344$                     &0.7854&1.10&1.9615&-0.0289\\
      &                             &      &1.15&2.1070& 0.0869\\
  \hline
\end{tabular}
\medskip

Columns 1 to 6 give the values of $x_{*}(2)$ where the Class I
and Class II twin shock solutions cross the sonic critical line
the second time, $x_{*}(1)$ and $m_{0}$ for Class I solutions,
$D$ for Class II solutions, the first shock location $x_s(1)$,
the second shock location $x_{s}(2)$, the mass density parameter
$A$ and the speed parameter $V$.
\end{minipage}
\end{table*}

\subsection[]{Novel Class III MHD Shock Solutions }

\begin{figure}
 \includegraphics[width=3in]{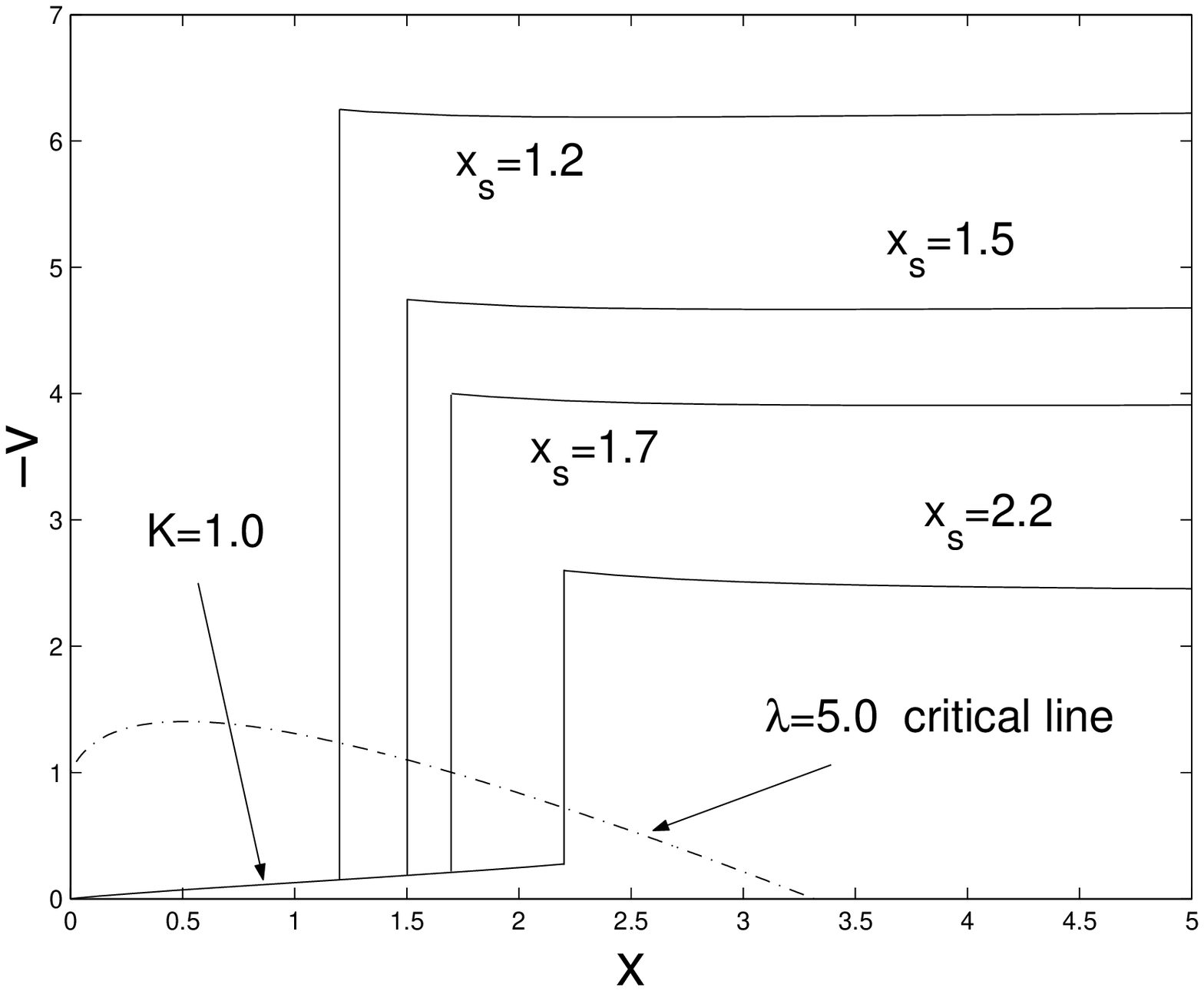}
 \caption{\label{newclass}
 Semi-complete Class III MHD shock solutions in terms of
 $-v(x)$ versus $x$ with $\lambda=5.0$ and $K=1.0$ for
 shock location $x_{s}=1.2,\ 1.5,\ 1.7,\ 2.2$, respectively.
 Asymptotic MHD solutions (iv) given by equations
 (\ref{shockclassIIIv}) and (\ref{shockclassIIIa}) are
 specified near the origin. The dash-dotted line represents
 the magnetosonic critical curve. The speed ratio $v_A/a$
 versus $x$ is shown in Fig. \ref{newclassalfven}. }
\end{figure}

\begin{figure}
 \includegraphics[width=3in]{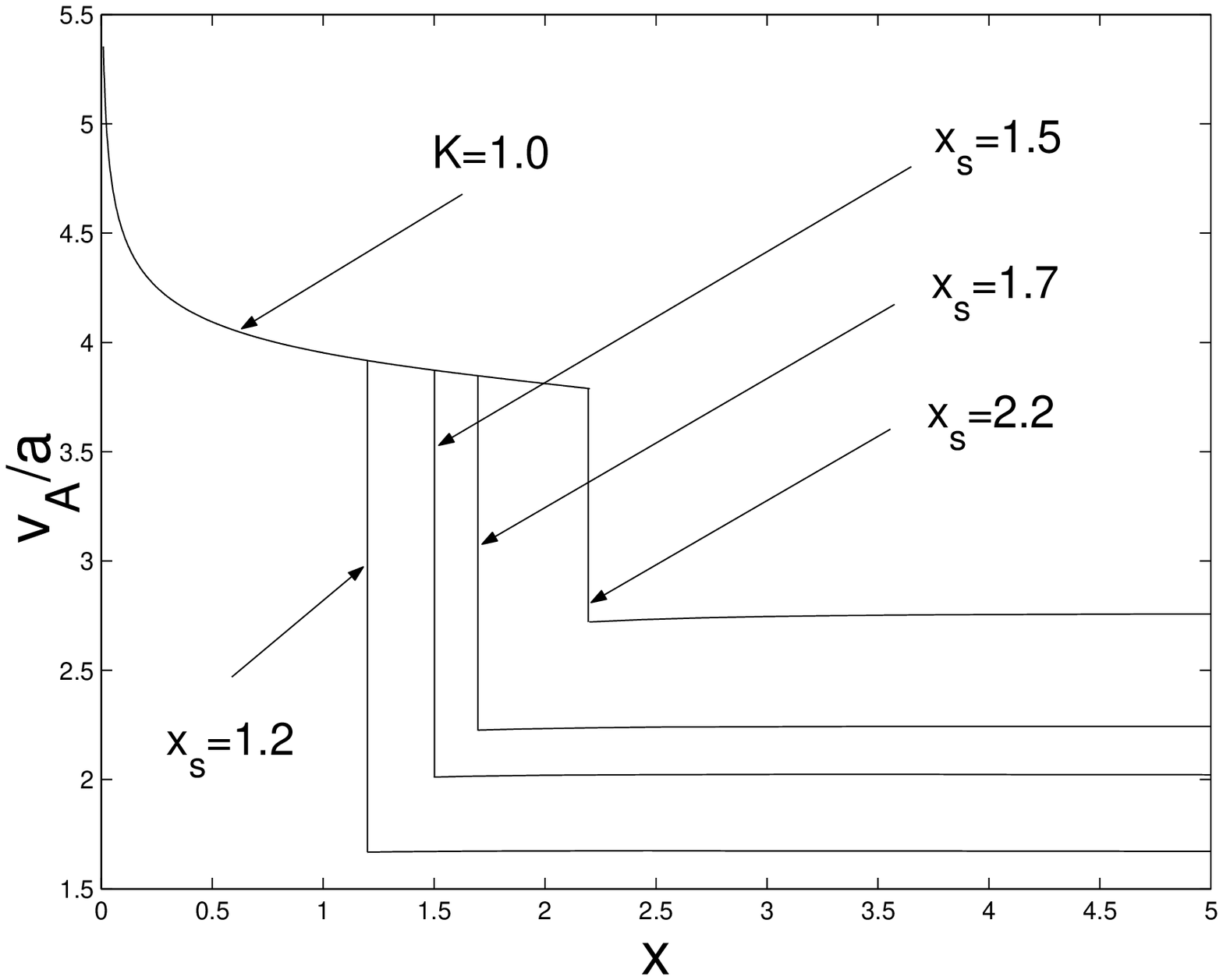}
 \caption{\label{newclassalfven}
 The ratio of the Alfv\'en speed to the isothermal sound
 speed $v_A/a$ versus $x$ with $\lambda=5.0$ and $K=1.0$,
 corresponding to the MHD shock results in Fig. \ref{newclass}.
 Class III MHD shock solutions with the MHD shock location at
 $x_{s}=1.2,\ 1.5,\ 1.7,\ 2.2$, respectively.  }
\end{figure}

In our numerical investigation, we also construct a unique Class III
solutions exclusive to MHD shock solutions. This type of solutions
carry a mixed feature of both Class I and Class II solutions. Their
reduced velocity is a finite convergent inflow, similar to Class II
solutions towards the centre. Their reduced mass density is divergent,
characteristic of Class I solutions towards the centre. Starting from
equations (\ref{shockclassIIIv}) and (\ref{shockclassIIIa}), we can
construct these new Class III MHD solutions. In order to assure the
existence of this Class III solutions, $\lambda$ should be greater than
4 (Wang \& Lou 2006). As examples of illustration, we take $K=1.0$,
$\lambda=5.0$, $H=[2-\lambda+(\lambda^2-4\lambda)^{1/2}]/2$ and
$x_{s}=1.2,\ 1.5,\ 1.7,\ 2.2$, respectively. In Figs. \ref{newclass}
and \ref{newclassalfven}, we present these Class III MHD shock solutions.

\subsection[]{Two-Temperature Similarity MHD Shocks }

We have only studied so far the case of $\tau=1$ in the preceding
sections. This means that both the downstream and upstream
magnetofluids across an MHD shock have the same thermal
temperature, i.e., the entire magnetofluid has same thermal sound
speed $a_{u}=a_{d}$. In some astrophysical systems (e.g., HII
regions around luminous massive OB stars), the factor of determining
the ionization temperature, mean atomic mass, etc., would vary from
an interior region to an exterior region. In this section, we shall
discuss a magnetized gas medium of two different temperatures across
an MHD shock. In our following consideration, the downstream sound
speed $a_{d}$ is thus no longer equal to the upstream sound speed
$a_{u}$. In most astrophysical systems, the downstream sound speed
should be faster than the upstream sound speed, while the downstream
and upstream fluids can be regarded as isothermal fluids separately.
With this in mind, we assume $a_{d}/a_{u}>1$ in our following MHD
model analysis.

Let us begin with the MHD shock solution for the case of $V=0$ at
$x\rightarrow+\infty$. Here, we take Class I MHD solutions with a
magnetosonic critical point $x_{*}(1)=0.103$ and Class II MHD
solutions with a central reduced density parameter $D=0.1$ as
examples of illustration. For different values of sound speed
ratio $a_{d}/a_{u}$, we obtain the Class I and Class II MHD
shock solutions through matching the upstream and downstream
solutions in the speed-density phase diagram.

\begin{figure}
 \includegraphics[width=3in]{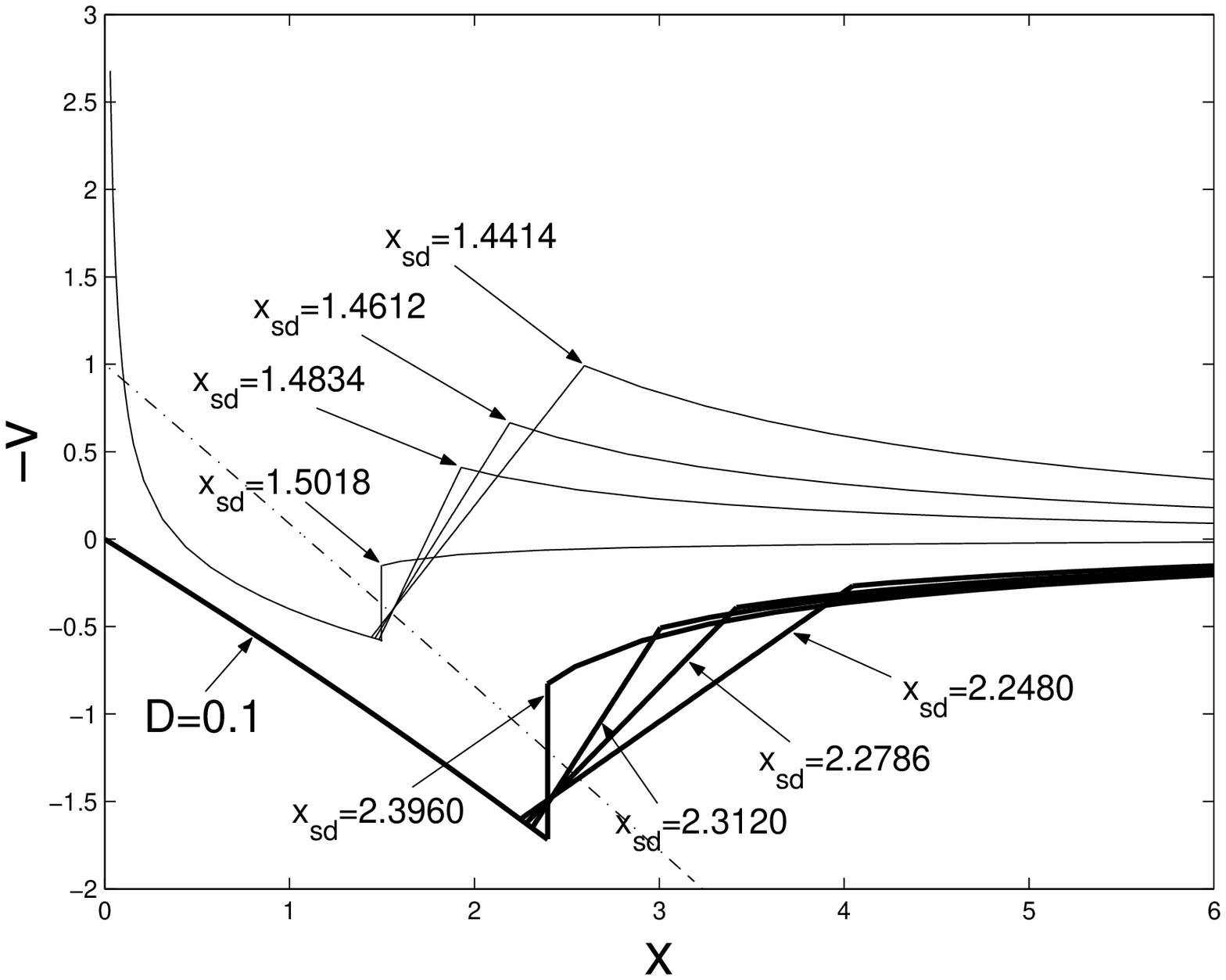}
 \caption{\label{twotempbothclass}
 Class I (light solid curves) two-temperature MHD shock
 breeze solutions with $\lambda=0.1$, $x_{*}=0.103$ and
 different density ratio $\alpha_{d}/\alpha_{u}\geq1$.
 Class II (heavy solid curves) two-temperature MHD shock breeze
 solutions with $\lambda=0.1$, $D=0.1$ and different density
 ratio $\alpha_{d}/\alpha_{u}\geq1$. The dash-dotted line
 represents the magnetosonic critical curve. The reduced
 transverse magnetic field strength $b(x)$ versus $x$ is
 shown in Fig. \ref{bfieldtwotempbothclass}. }
\end{figure}

\begin{figure}
 \includegraphics[width=3in]{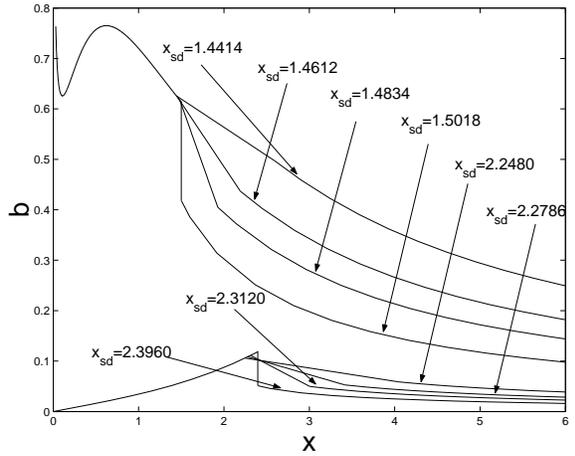}
 \caption{\label{bfieldtwotempbothclass}
 Reduced transverse magnetic field strength $b(x)$ versus $x$
 corresponding to the results in Fig. \ref{twotempbothclass}.
 Class I two-temperature MHD shock breeze solutions with
 $\lambda=0.1$, $x_{*}=0.103$ and different density ratio
 $\alpha_{d}/\alpha_{u}\geq1$. Class II two-temperature
 MHD shock breeze solutions with $\lambda=0.1$, $D=0.1$
 and different density ratio $\alpha_{d}/\alpha_{u}\geq1$.  }
\end{figure}

In Fig. \ref{twotempbothclass} and Fig.
\ref{bfieldtwotempbothclass}, we display both classes of MHD shock
solutions with their parameters summarized in Table 5.
In both classes of MHD shock solutions, as the ratio $a_{d}/a_{u}$
increases, the downstream shock location $x_{sd}$, namely, the
Mach number of the downstream shock becomes smaller. The mass
parameter $A$ increases with an increasing $a_{d}/a_{u}$ ratio.
When the mass parameter $A$ is less than 2, the upstream solution
is an MHD outflow breeze. The MHD breeze is stronger with a smaller
value of $A$. In the case of the mass parameter $A$ larger than 2,
the inflow speed becomes weaker when $A$ becomes smaller.

\begin{figure}
 \includegraphics[width=3in]{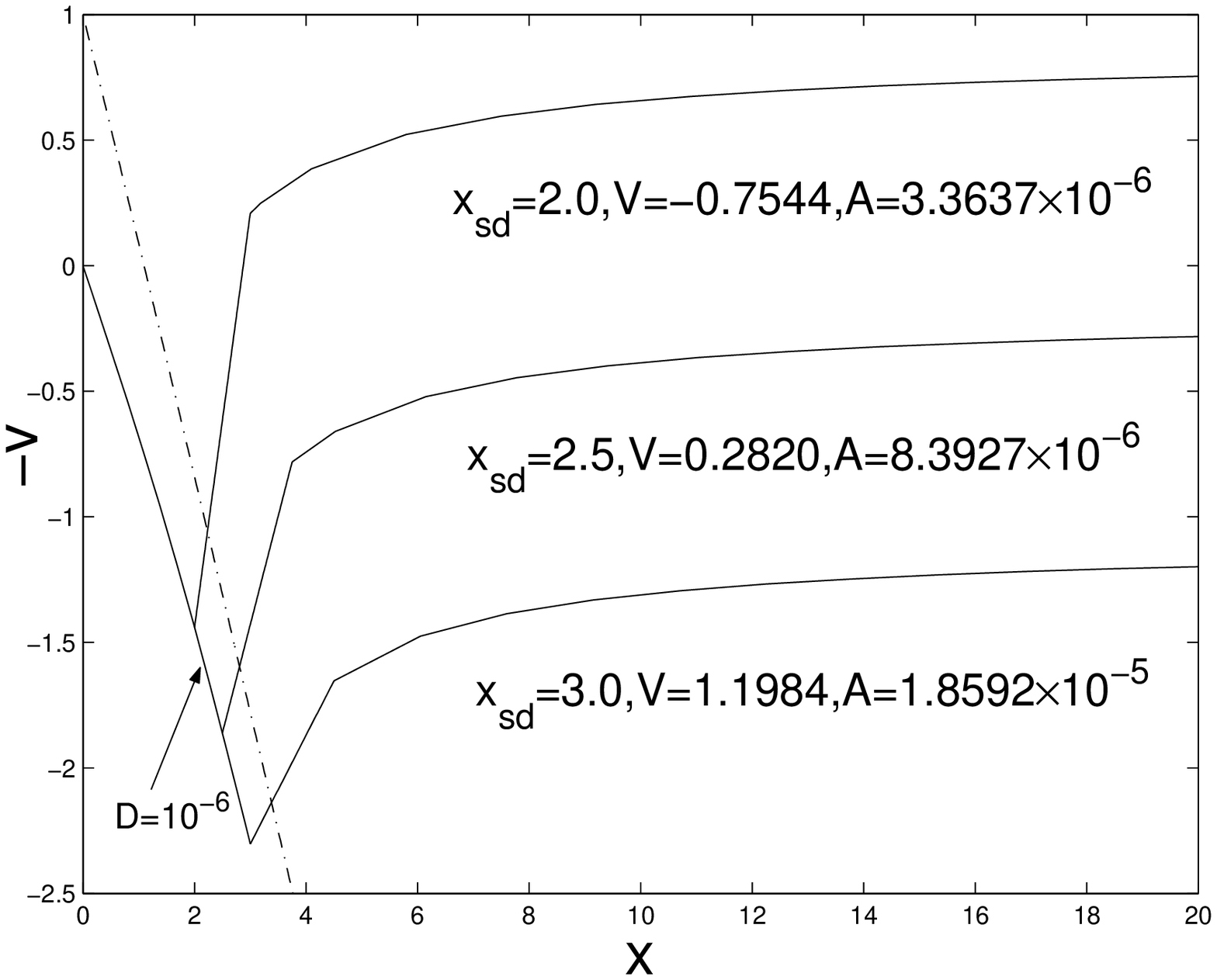}
 \caption{\label{twotempbtiny} Class II MHD shock solutions with
 $\lambda=0.1$ and $D=10^{-6}$ shown in terms of $-v(x)$ versus $x$ by light
 solid curves. The dash-dotted line represents the magnetosonic critical
 curve. The reduced transverse magnetic field $b(x)$ versus $x$ is shown
 in Fig. \ref{bfieldtwotempbtiny}.}
\end{figure}

\begin{figure}
 \includegraphics[width=3in]{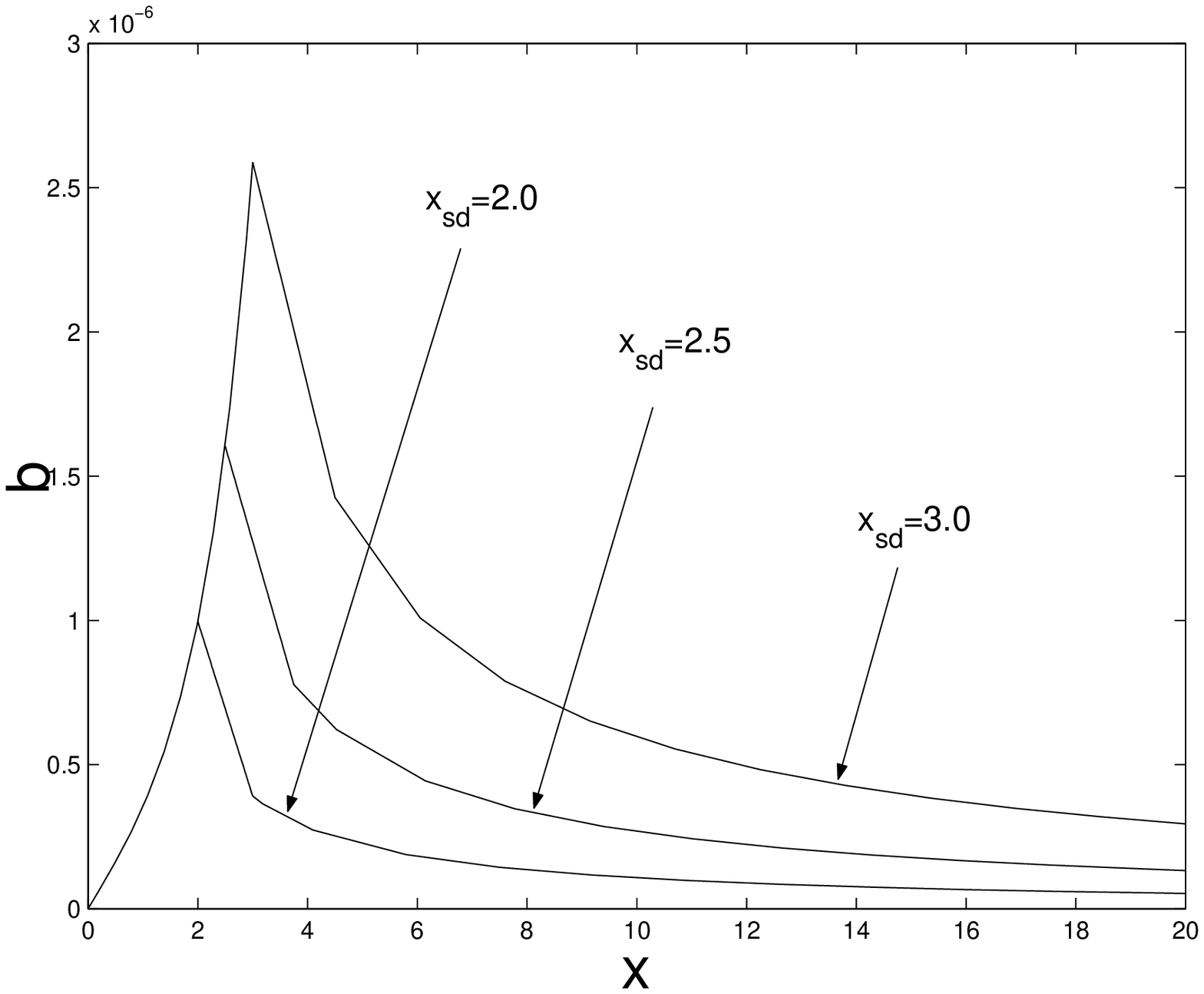}
 \caption{\label{bfieldtwotempbtiny} The reduced transverse magnetic field
 $b(x)$ versus $x$ corresponding to the Class II MHD shock solutions with
 $\lambda=0.1$ and $D=10^{-6}$ shown in Fig. \ref{twotempbtiny}. }
\end{figure}

We now investigate the case for the velocity parameter $V\neq 0$.
For instance, we choose $a_{d}/a_{u}=1.5$ to study Class II MHD
shock solutions. Different locations of $x_{sd}$ lead to different
MHD shock solutions. In Fig. \ref{twotempbtiny} and Fig.
\ref{bfieldtwotempbtiny}, we present different MHD shock solutions
with the reduced central density parameter $D=10^{-6}$.

%

\begin{table*}
\centering
\begin{minipage}{115mm}
\caption{Two-temperature Class I and II similarity MHD shock breeze
solutions with different values of sound speed ratio $a_{d}/a_{u}$.}
\begin{tabular}{|l|l|l|l|l|l|l|l|l|}
  \hline
  \hline
  Parameter &$a_{d}/a_{u}$&$A$&$x_{sd}$&$x_{su}$
  &$v_{d}$&$\alpha_{d}$&$v_{u}$&$\alpha_{u}$\\
  \hline
  $x_{*}(1)=0.103$&1  &1.8530&1.5018&1.5018&0.5805&1.2907&0.1507 &0.8802\\
                  &1.3&2.7594&1.4834&1.9284&0.5750&1.3150&-0.4102&0.6640\\
                  &1.5&3.5226&1.4612&2.1918&0.5682&1.3452&-0.6657&0.6305\\
                  &1.8&4.9048&1.4414&2.5945&0.5622&1.3729&-0.9916&0.6059\\
  \hline
  $D=0.1$         &1  &0.2976&2.3960&2.3960&1.7176&0.1563&0.8273 &0.0676\\
                  &1.3&0.4226&2.3120&3.0056&1.6517&0.1521&0.5082 &0.0523\\
                  &1.5&0.5288&2.2786&3.4179&1.6256&0.1505&0.3913 &0.0487\\
                  &1.8&0.7206&2.2480&4.0464&1.6018&0.1491&0.2675 &0.0459\\
  \hline
\end{tabular}
\medskip

Columns 1 to 9 contains relevant parameters of MHD shock
solutions: $x_{*}(1)$ for the magnetosonic critical point of Class
I MHD solutions and $D$ for the central reduced density parameter
of Class II MHD solutions; the sound speed ratio of the downstream
$a_{d}$ to the upstream $a_{u}$; the upstream mass density
parameter $A$; $x_{sd}$ is the shock location in terms of
downstream $a_d$; $x_{su}$ is the shock location in terms of
upstream $a_u$; the downstream reduced speed $v_{d}$; the
downstream reduced density $\alpha_{d}$; the upstream reduced
speed $v_{u}$; the upstream reduced density $\alpha_{u}$.
\end{minipage}
\end{table*}

\begin{table*}
\centering
\begin{minipage}{115mm}
\caption{Two-temperature Class II MHD shock solutions for
sound speed ratio $a_{d}/a_{u}=1.5$ with $D=10^{-6}$. }
\begin{tabular}{|c|c|c|c|c|c|c|c|}
  \hline
  \hline
  $x_{sd}$ & $x_{su}$ & $v_{d}$ & $\alpha_{d}$ & $v_{u}$ &
  $\alpha_{u}$ & $A$ & $V$ \\
  \hline
  2.0&3.00&1.4407&$1.5753\times10^{-6}$&-0.2092&$4.1184\times10^{-7}$&
  $3.3637\times10^{-6}$&-0.7544\\
  2.5&3.75&1.8608&$2.0284\times10^{-6}$&0.7814 &$6.5518\times10^{-7}$&
  $8.3927\times10^{-6}$&0.2820\\
  3.0&4.50&2.3032&$2.7282\times10^{-6}$&1.6533 &$1.0016\times10^{-6}$&
  $1.8592\times10^{-5}$&1.1984\\
  \hline
\end{tabular}
Columns 1 to 8 summarize relevant parameters for MHD shock
solutions. $x_{sd}$ is the shock location in terms of downstream
$a_d$; $x_{su}$ is the shock location in terms of upstream $a_u$;
the downstream reduced velocity is $v_{d}$; the downstream reduced
density is $\alpha_{d}$; the upstream reduced velocity is $v_{u}$;
the upstream reduced density is $\alpha_{u}$; $A$ is for the mass
density parameter of upstream and $V$ is the upstream speed
parameter.
\end{minipage}
\end{table*}

\section{Notes and Discussion}

%
%
%
%
%
%
%
%
In this paper, we investigated quasi-spherical self-similar shock
solutions for magnetized self-gravitating isothermal fluids in the
semi-complete solution space. The relevant similarity MHD shock
solutions are obtained and classified by comparing with results of
previous analyses
\citep{th95,shu02,loushen,shenlou04,bianlou05,yulou05}. For a
closely relevant similarity analysis on a magnetized polytropic
gas, the reader is referred to Wang \& Lou (2006). Such mEECC
similarity solutions, which depict an expanding envelope with a
concurrent collapsing core, are derived from nonlinear MHD
equations and may be applied to various astrophysical MHD
processes with appropriate adaptations.


As noted by Yu \& Lou (2005), the velocity and density profiles of
an EECC shock model agree better with observations of cloud B335
than the results of the EWCS model. The magnetized EECC shock
solutions qualitatively retain this essential feature of
nonmagnetized EECC solutions (Shen \& Lou 2004). The inclusion of
a random magnetic field is more realistic and can be further
utilized to model diagnostic features such as synchrotron
emissions etc. Due to current observational limits, no empirical
magnetic field information is immediately available for the most
inner core of cloud system B335 (e.g., Wolf et al. 2003). Our MHD
EECC shock solutions may give a rough estimate for the magnetic
field strength in the most inner core of star forming cloud B335
as an example. For the starless cloud system B335, we may take an
infalling region of a spatial scale $\sim 1.5\times10^{4}$ AU, a
cloud mass of $\sim 4M_{\odot}$, a random magnetic field strength
of $\sim 134\mu$G (e.g., Wolf et al. 2003)
and a thermal sound speed $a\sim 0.23\hbox{ km s}^{-1}$. For a
quasi-spherical accretion MHD shock to exist within such a cloud
system, the core magnetic field strength may reach $\sim 1$mG,
which is several times strong than our previous estimate of $\sim
300\mu$G, depending on the strength of such an MHD shock.
Observations indicate that some T Tauri stars may have surface
magnetic field strengths of the order of $10^{3}$G, which is much
greater than our estimate. Possibly, our quasi-spherical solutions
may not be valid when the spatial scale involved becomes much
smaller than $\sim 100$ AU; within such a scale, a circumstellar
disc as well as the stellar dynamo process would dominate among
various processes to amplify the protostellar magnetic fields
(n.b., a small-scale circumstellar disc and a relevant bipolar
outflow would be regarded as additional features in our
quasi-spherical mEECC shock scenario).


Such mEECC similarity solutions with a random magnetic field may
also be applied to the asymptotic giant branch (AGB) phase or
post-AGB phase in the late evolution stage of a low-mass
main-sequence star before the gradual emergence of a planetary
nebula (PN) system with a central magnetized white dwarf of a high
surface temperature in the range of $1\times 10^5\sim 2\times
10^5$K. The timescale of this evolution `gap' between these two
phases is estimated to be $\sim 10^3$yrs. Planetary nebulae and
proto-planetary nebulae (pPNs) are believed to be the ultimate
evolutionary stages of low- and intermediate-mass stars
(M$\leq8M_{\odot}$). PNs and pPNs appear on the sky as expanding
plasma clouds surrounding a luminous central star. Such expanding
clouds have a typical asymptotic speed of $\sim$ $10-20$ $\hbox{km
s}^{-1}$; the relevant mass loss rates fall within the range of
$\sim 10^{-8}$ to $\sim 10^{-4}M_{\odot}\hbox{ yr}^{-1}$.  With an
insufficient nuclear fuel supply from a certain epoch on, the
central region starts to contract and collapse while the outer
envelope continues to expand into a massive wind; it is highly
likely that the central collapse is accompanied by an outward
energetic surge to chase the slowly expanding envelope. Thus a
slow, dense stellar wind expelled during the AGB phase is followed
by a fast, tenuous magnetized wind driven off the collapsing
proto-white dwarf during the PN phase. The collision of magnetized
winds with two different flow speeds would inevitably generate an
MHD shock. In this process, the outer expansion removes stellar
envelope mass companied by a proto white dwarf produced at the
centre by central infall and collapse. Sufficiently far away from
the initial and boundary conditions, the system may gradually
evolve into an MHD phase representable by an MHD EECC similarity
solution during a timescale of a few hundred to several thousand
years. As noted out by Bian \& Lou (2005), the dimensionless mass
accretion rate $m_{0}$ should be in the range of $\sim
10^{-3}-10^{-4}$. In a PN, the dimensionless magnetic parameter
$\lambda$ is about $0.003$. Our Class I twin MHD shock solutions
appears to have the desired reduced mass accretion rate towards
the collapsed core.

Observations grossly indicate a magnetic field strength of $\sim
1$G near the stellar surface at $r\sim 1$AU during the AGB phase,
decreasing approximately to $\sim 10^{-3}$G at $r\sim 10^3$AU. In
the quasi-spherical self-similar MHD expansion regime of our model
analysis, the transverse magnetic field strength scales as
$B_{\perp}\propto r^{-1}$ (e.g., Lou 1993, 1994); for an expanding
MHD shock travelling outwards, the magnetic field downstream would
be enhanced by the existence of such an MHD shock.
Our model estimates are roughly consistent with the measured
magnetic field variations observed for AGB stars. The magnetic
field strength associated with an MHD EECC solution in the
innermost collapse region scales as $B\propto r^{-1/2}$. If we
take the strength of a surface magnetic field as $\sim 1$G at
$r\sim 1$AU and the radius of a proto-white dwarf as $\sim 6000$
km, then the magnetic field strength estimated at the surface of
a proto-white dwarf is roughly $10^{3}$G. An accretion MHD shock
would further increase this inner downstream magnetic field. This
enhancement is determined by the MHD shock strength. Much stronger
surface magnetic fields of a magnetic white dwarf ($\sim 10^{6}
-10^{9}$G) might be generated and sustained by intrinsic stellar
MHD dynamo processes driven by the convective differential
rotation inside an AGB star (e.g., Blackman et al. 2001).

We therefore hypothesize that dynamical evolution of an mEECC phase
of around or less than a few thousand years may be the missing
linkage between the AGB or post-AGB phase and the gradual appearance
of a PN. Depending on physical parameters of the low-mass progenitor
star, it may also happen that the degenerate CO core collapse and
subsequent material infalls during an mEECC phase lead to an eventual
core mass exceeding the Chandrasekhar mass limit of $1.4M_{\odot}$
and thus induce a supernova with an intensity determined by the
actual central mass accretion rate.

For the Crab Nebula as an example of magnetized SNR, we model the
nebular magnetic field as $B_{\perp}\propto r^{-1}$ during the outer
envelope expansion phase, ignoring complex interactions between the
central magnetized relativistic pulsar wind and the inner nebula.
In the presence of shocks, as a flow passes through a shock from
downstream to upstream, the magnetic field strength would decrease
by a factor of a few. Thus, the magnetic field strength in the
flow would change in the presence of shocks as compared with mEECC
solutions without shocks, although not significantly. For a
neutron star of a radius $\sim 10$ km and a typical dipolar
magnetic field of strength $\sim 10^{10}$G  (we exclude for the
moment those unusual high-field magnetars, such as SGRs and AXPs
whose magnetic field strength may reach as high as $10^{14}\sim
10^{15}$G), the magnetic field would decrease to $\sim10^{-3}$G at
a distance of about several parsecs (Yu \& Lou 2005). When an MHD
shock is included and for the same neutron star magnetic field,
the outer nebular magnetic field would become weaker $\sim
10^{-4}$G at a distance of about several parsecs. These estimates
more or less agree with the magnetized envelope expansion portion
of our mEECC shock solutions. In addition to this order of
magnitude agreement with observations for magnetic field
strengths, mysterious central structures of the Cran Nebula like
wisps and knots are likely produced by reverse fast MHD shock
waves as a result of slightly inhomogeneous pulsar wind streams
emanating from the fast spinning pulsar (e.g., Lou 1998). These
quasi-stationary reverse MHD shocks in space can effectively
produce relativistic particles and synchrotron emissions from
relativistic electrons gyrating rapidly around magnetic field
lines surrounding these compact objects. The familiar triple ring
structure of SN1987A may be associated with MHD shocks (Tanaka \&
Washimi 2002).

Unlike the Crab Nebula, which is primarily powered by the
rotational energy through a relativistic pulsar wind, the slowly
rotating magnetars are thought to be powered by the decay of
extremely intense magnetic field. The light curves may thus
represent an adiabatically expanding population of electrons
accelerated at a particularly active phase, which could be
produced by the ejecta colliding with a pre-existing shell. Such a
shell is naturally made by SGR itself, because its quiescent wind
of a luminosity $\sim 10^{34}$ erg $\mathrm{ s}^{-1}$ will sweep
up a bow shock of a stand-off distance $\sim 10^{16}$ cm as it
moves through the interstellar medium (ISM) at a typical neutron
star velocity of $\sim 200$ km $\mathrm{ s}^{-1}$. If this
pre-existing shell is hit by giant flares from a SGR with an
energy of $\sim 10^{43}-10^{44}$ erg, MHD shocks will naturally
occur and be swept outward, resulting in a violent episode of
relativistic particle acceleration that deposits much of the
energy into a steadily expanding synchrotron-emitting shell after
giant flares. Polarized emissions from SGRs also indicate that
magnetic field should play a crucial role in understanding SGR
phenomena.

As massive OB stars turn on their core nuclear reactions in
magnetized molecular clouds, the interstellar medium of
surrounding neutral hydrogen gas strongly absorbs ultraviolet
radiations from OB stars and becomes ionized to form luminous HII
regions (e.g., Str$\mathrm{\ddot{o}}$mgren 1939; Osterbrock 1989;
Shu et al. 2002). As the ionization front sweeps through the
neutral cloud, magnetized gas flows driven by pressure gradients
develop between HII and HI regions as well as within HII regions,
and MHD shocks can naturally emerge in these processes.

For astrophsical systems of much larger scales such as clusters of
galaxies (e.g., Sarazin 1988; Fabian 1994),
the similarity MHD shock solutions may be valuable for
understanding a certain phase of their evolution involving
magnetic fields (e.g., Hu \& Lou 2004; Lou 2005). Chandra
observations show that the continuous blowing of bubbles by the
central radio source would lead to the propagation of shocks seen
as the observed fronts and ripples, gives a rate of working which
balances the radiative cooling within the cluster core (Fabian et
al. 2003). With appropriate adaptation, our MHD shock model may
offer interesting interpretations of such phenomena.

\section*{Acknowledgments}
This research has been supported in part by the ASCI Center for
Astrophysical Thermonuclear Flashes at the University of Chicago
under the Department of Energy contract B341495, by the Special
Funds for Major State Basic Science Research Projects of China, by
the Tsinghua Center for Astrophysics, by the Collaborative
Research Fund from the National Science Foundation of China (NSFC)
for Young Outstanding Overseas Chinese Scholars (NSFC 10028306) at
the National Astronomical Observatories of China, Chinese Academy
of Sciences, by the NSFC grants 10373009 and 10533020 at the
Tsinghua University, and by the SRFDP 20050003088
and the Yangtze Endowment from the Ministry of Education at the
Tsinghua University. Affiliated institutions of Y-QL share this
contribution.

\vskip 0.4cm


\appendix
\section[]{\ \ Derivation of MHD
Perpendicular Shock Condition}

We have the following basic definitions and relations
\[
\frac{\rho_{2}}{\rho_{1}}=X,\quad \frac{u_{2}}{u_{1}}=\frac{1}{X},
\quad \frac{p_{2}}{p_{1}}=X, \quad \frac{B_{2}}{B_{1}}=X, \quad
M_{1}=\frac{u_{1}}{a}\ .
\]
Momentum conservation (\ref{onetempmomentum}) can be written as
\[
X p_{1}+X^{2}B_{1}^{2}/(8\pi) +\rho_{1}u_{1}^{2}/X
=p_{1}+B_{1}^{2}/(8\pi) +\rho_{1}u_{1}^{2}\ .
\]
%
%
Using the following two relations
\[
\rho_{1}u_{1}^{2}=\rho_{1}a^{2}
\frac{u_{1}^{2}}{a^{2}}=p_{1}M_{1}^{2},
\qquad\qquad
\frac{B_{1}^{2}}{8\pi}=\frac{p_{1}}{\beta_{1}}\ ,
\]
%
%
the foregoing momentum equation becomes
\[
Xp_{1}+X^{2}p_{1}/\beta_{1}+p_{1}M_{1}^{2}/X=
p_{1}+p_{1}/\beta_{1}+p_{1}M_{1}^{2}\ ,
\]
or simply
\[
X+X^{2}/\beta_{1}+M_{1}^{2}/X
=1+1/\beta_{1}+M_{1}^{2}\ .
\]
It then follows immediately that
\[
(X-1)+
(X-1)(X+1)/\beta_{1}=M_{1}^{2}(X-1)/X\ .
\]
The degenerate case of $X-1=0$ would be trivial.
Removing this factor $X-1$, we readily arrive at
\[
1+(X+1)/\beta_{1}=M_{1}^{2}/X\ .
\]
This is the quadratic equation governing the
density ratio $X$, exactly the same as equation
(\ref{onetempf}) in the main text.

\label{lastpage}
\end{document}